\documentclass[reprint,prd,nofootinbib,preprintnumbers]{revtex4}
\pdfoutput=1
\usepackage{graphicx,epsfig,fancyvrb,makeidx,latexsym,amssymb,amsmath,color}
\def\lsim{\mathrel{\rlap{\lower4pt\hbox{\hskip1pt$\sim$}}
    \raise1pt\hbox{$<$}}}    
\def\gsim{\mathrel{\rlap{\lower4pt\hbox{\hskip1pt$\sim$}}
    \raise1pt\hbox{$>$}}}                

\newcommand{\beq}{\begin{eqnarray}}
\newcommand{\eeq}{\end{eqnarray}}

\begin{document}
\preprint{MAN/HEP/2013/13}
\title  {Neutrino Mass and Dark Matter in light of recent AMS-02 results}
\author{P. S. Bhupal Dev,$^{a}$ Dilip Kumar Ghosh,$^{b}$ Nobuchika Okada$^{c}$ and Ipsita Saha$^{b}$}
\affiliation{\vspace{2mm}
$^a$ Consortium for Fundamental Physics, School of Physics and Astronomy, University of Manchester, Manchester, M13 9PL, United Kingdom.\\
$^b$ Department of Theoretical Physics,
Indian Association for the Cultivation of Science,
2A \& 2B Raja S.C. Mullick Road, Kolkata 700032, India.\\
$^c$ Department of Physics and Astronomy, University of Alabama, Tuscaloosa, AL 35487, USA.}
\begin{abstract}
\vspace{0.2cm}
\centerline{\bf ABSTRACT}
\vspace{1mm}
We study a simple extension of the Standard Model supplemented by an 
electroweak triplet scalar field to accommodate small neutrino masses by the 
type-II seesaw mechanism, while an additional singlet scalar field can play the role 
of cold dark matter (DM) in our Universe. This DM candidate is leptophilic for a wide range of model parameter space, and the lepton flux due to its annihilation carries information about the neutrino mass hierarchy. Using the recently released high precision data on positron fraction and flux from the AMS-02 experiment, we examine the DM interpretation of the observed positron excess in our model for two kinematically distinct scenarios with the DM and triplet scalar masses (a) non-degenerate ($m_{\rm DM}\gg m_{\Delta}$), and (b) quasi-degenerate ($m_{\rm DM} \simeq m_\Delta$). We find that a good fit to the AMS-02 data can be obtained in both cases (a) and (b) with a normal hierarchy of neutrino masses, while the inverted hierarchy case is somewhat disfavored. Although we require a larger boost factor for the normal hierarchy case, this is still consistent 
with the current upper limits derived from Fermi-LAT and IceCube data for case (a). 
Moreover, the absence of an excess anti-proton flux as suggested by PAMELA data sets an indirect 
upper limit on the DM-nucleon spin-independent elastic 
scattering cross section which is stronger than the existing 
DM direct detection bound from LUX in the AMS-02 preferred DM mass range.  
\end{abstract}
\maketitle
\section{Introduction}\label{sec1}
Recently the AMS-02 collaboration has reported a significant excess in the 
cosmic-ray positron fraction over the expected background in the energy range 
10 - 350 GeV~\cite{AMS, AMS-update}. This high precision measurement of the positron 
fraction confirms its steady rise in this energy range as previously observed by  PAMELA~\cite{pamela1, pamela2} (recently updated in~\cite{pamela-update}) and Fermi-LAT~\cite{fermi} (for earlier experiments, see e.g.,  TS93~\cite{ts93}, WIZARDS/CAPRICE~\cite{wizard}, HEAT~\cite{heat1, heat2} and AMS-01~\cite{ams1}). Such a rise cannot be explained in the standard astrophysical picture 
where cosmic-ray positrons and anti-protons are mostly produced as secondary particles due to nuclear interactions of high energy cosmic-ray protons with the interstellar medium (ISM), and no corresponding increase in the anti-proton flux has been observed~\cite{aph, ap1, ap2}. Moreover, the observed energy spectrum of positron fraction shows no fine structure, and the positron to electron ratio shows no observable anisotropy~\cite{AMS-update} (which would have been induced by primary 
sources of cosmic-ray positrons and electrons). 
These observations, together with the excess in total electron/positron flux observed  by  ATIC~\cite{atic} (also PPB-BETS~\cite{bets}), HESS~\cite{hess1, hess2}, Fermi-LAT~\cite{fermi, fermi-flux1, fermi-flux2}, PAMELA~\cite{pamela-flux}, 
and more recently by AMS-02~\cite{AMS-update} in the energy range 1 - 700 GeV, suggest the existence of some 
new relatively local 
primary source(s) of high energy electrons/positrons, from either a particle physics or an astrophysical origin (for reviews, see e.g.,~\cite{review1, review1a}).\footnote{Yet another possibility to explain the excess positron fraction without invoking a new primary source is to modify the galactic cosmic-ray propagation models (see e.g.,~\cite{prop1, prop2, prop3}).}

A possible particle physics interpretation of the positron excess observed by AMS-02 is the annihilation/decay 
of TeV-scale dark matter (DM) particle(s)~\cite{DM1, DM2, DM3, DM4, DM5, DM6, DM7, DM8, DM9, DM10, yuan, hooper0}.\footnote{For earlier studies in the context of PAMELA and Fermi-LAT results, see e.g.~\cite{review1a, review2} and references therein.} This scenario usually requires much 
larger annihilation cross section than that required to explain the DM relic abundance in the Universe through thermal freeze-out. In addition, since the PAMELA anti-proton flux shows no excess over expected cosmic-ray background~\cite{ap1, ap2}, any DM model attempting to explain the positron excess should be leptophilic, i.e., predominantly producing leptonic final states rather than hadrons. These models are also subject to strong constraints due to radio flux via synchrotron 
emission~\cite{bertone, bell} and diffuse $\gamma$-ray flux via bremsstrahlung and inverse Compton scattering~\cite{bertone, bell, Cirelli:2009dv, Papucci:2009gd, Ackermann:2011wa, Hooper:2012sr}. Besides these cosmic-ray signals, a large DM annihilation rate could also affect the light element 
abundances in the Big Bang Nucleosynthesis (BBN) epoch~\cite{Jedamzik:2006xz, 
Hisano:2009rc}, and also affect the ionization history, thereby distorting the Cosmic Microwave Background (CMB) spectrum~\cite{Slatyer:2009yq, Huetsi:2009ex, Cirelli:2009bb, Galli:2011rz, Cline:2013fm}. Using the precision 
positron excess data from AMS-02, stringent upper limits on the DM annihilation rate 
have been derived~\cite{kopp, hooper} for DM masses below $\sim 300$ GeV, annihilating to leptonic 
final states.        

Astrophysical accelerators like pulsars (fast rotating magnetized neutron stars) and pulsar wind nebulae are also known to be powerful sources of energetic electron-positron pairs within our Galaxy, and provide an alternative explanation of the AMS-02 result~\cite{yuan, hooper0, pulsar1, pulsar2}, though plagued by large inherent astrophysical uncertainties, e.g., the number and energy distribution of electron-positron pairs injected into the 
ISM. Although currently we cannot discriminate between a DM and pulsar origin of the observed positron excess, the future precision data from AMS-02, especially on secondary-to-primary ratios of cosmic-ray nuclei species, may be able to solve this longstanding issue~\cite{pulsar-DM1, pulsar-DM2, pulsar-DM3, pulsar-DM4}. Note that that if the observed positron excess has a particle physics origin, it is expected to be isotropic in nature. The recent updates from 
AMS-02~\cite{AMS-update} indicate such an isotropy in the fluctuation of the positron ratio, 
with an upper limit on the amplitude of dipole anisotropy, $\delta\leq 0.030$ at 95\% CL for any axis in galactic coordinates in the energy range of 16 - 350 GeV.  However, this cannot yet be taken as an evidence against an astrophysical origin of the positron excess due to the inherent uncertainties associated with source and propagation 
effects (see e.g.,~\cite{pulsar-DM4}). 

In this paper we provide a particle physics interpretation of the AMS-02 results in a simple extension of the SM accommodating non-zero neutrino masses as well as a cold DM candidate. The particle content of the SM is supplemented by an $SU(2)_L$-triplet scalar field $\mathbf{\Delta}\equiv (\Delta^{++},\Delta^+,\Delta^0)$, thus leading to non-zero neutrino masses via type-II seesaw 
mechanism~\cite{type2a, type2b, type2c, type2d, type2e}. Apart from explaining neutrino masses, some other nice features of 
type-II seesaw models are: (a) restoration of the electroweak vacuum stability up to the Planck scale for a large range of model parameter space~\cite{vacuum1, vacuum2, vacuum3, vacuum4, vacuum5, vacuum6, DGOS, vacuum7}, and 
(b) rich LHC phenomenology for a low seesaw scale ~\cite{LHC1, LHC2, LHC3, goran} since the $SU(2)_L$-triplet scalars directly couple to the SM gauge bosons and leptons. In addition, light triplet scalars can 
significantly modify the Higgs-to-diphoton~\cite{goran, hgg1, hgg2} and Higgs-to-$Z$+photon~\cite{hzg1, hzg2, hzg3} 
decay rates which 
are correlated in the type-II seesaw model for most of the allowed 
parameter space~\cite{DGOS, Chen:2013dh}.  In fact, for an observable ($\gsim 10\%$) enhancement in 
the Higgs-to-diphoton rate  over its SM expectation, there exists an upper bound on the triplet mass ($\lsim 450$ GeV) from vacuum stability requirements~\cite{DGOS}, thus making these triplet scalars completely accessible at the LHC. It may be noted here that the latest ATLAS results still show a persistent excess in the $h\to \gamma\gamma$ signal strength:   $\hat{\mu}_{\gamma\gamma}^{\rm ATLAS}=1.55^{+0.33}_{-0.28}$~\cite{atlas-hgg}, whereas the corresponding CMS best-fit value is much lower: $\hat{\mu}_{\gamma\gamma}^{\rm CMS, MVA}=0.78^{+0.28}_{-0.26}$ from 
MVA analysis, whereas their cut-based analysis gives $\hat{\mu}_{\gamma\gamma}^{\rm CMS, cut}=1.11^{+0.32}_{-0.30}$~\cite{cms-hgg}. More data on precision Higgs measurements in future should be able to resolve this discrepancy between the two experiments. 

The required DM content of the Universe~\cite{Planck} can be easily 
accommodated in a non-supersymmetric scenario by adding a SM singlet scalar 
field ($D$)~\cite{singlet1, singlet2, singlet3, singlet4, singlet5} whose stability can be guaranteed by assigning it an odd $Z_2$-parity. When embedded in the type-II seesaw, the leptophilic 
nature of $D$, as required to fit the AMS-02 data, can be attributed to its 
interactions with the $SU(2)_L$-triplet scalar field $\mathbf{\Delta}$ which dominantly decays to leptonic final states for a small triplet vacuum expectation value (VEV) $v_\Delta<0.1$ MeV. An important distinguishing feature of this model is that the final-state lepton flavor due to DM annihilation is 
intimately connected to the neutrino mass hierarchy through the 
Dirac Yukawa coupling in the model. Thus the amount of electron/positron flux produced from the DM annihilation, and hence, the goodness of fit to the AMS-02 data, which now offer an unprecedented accuracy, can provide a new probe of the neutrino mass hierarchy in this class of models.\footnote{For some 
earlier analyses explaining non-zero neutrino masses as well as the positron excess in the context of PAMELA results, see e.g.,~\cite{Gogoladze:2009gi, zhang, Chen:2009ew, Bajc:2010qj}. For another class of naturally leptophilic DM scenarios, see~\cite{khlopov1, khlopov2, khlopov3}.}  

In its minimal version, this model requires a large ``boost factor" of order $10^3$ - $10^4$ in the DM annihilation rate to explain the observed positron excess. Such a boost factor could  
have either an astrophysical origin due to small-scale inhomogeneities in the DM density distribution which cannot be excluded even with the highest resolution numerical simulations available at present~\cite{Kuhlen:2012ft}, or a particle physics origin due to the Breit-Wigner enhancement mechanism~\cite{bw1, bw2, bw3} which can be easily implemented in our model, for instance, by introducing another $Z_2$-even singlet real scalar field $S$ coupling to both $D$ and $\mathbf{\Delta}$~\cite{Gogoladze:2009gi}.
We find that such large boost factors, as required to explain the AMS-02 positron data, are still consistent with the current upper limits on DM annihilation to neutrinos and gamma-rays derived from 
IceCube~\cite{ic1,ic2} and Fermi-LAT~\cite{Ackermann:2013yva} data respectively, except for the case in which the DM and triplet masses are quasi-degenerate. We perform a $\chi^2$-minimization taking into account the AMS-02 positron fraction data to find the $3\sigma$ allowed ranges of the DM mass and boost factor in our model. 
It turns out that the goodness of fit for the normal hierarchy (NH) case is somewhat better than for the inverted hierarchy (IH) case, thus implying that this model prefers a NH of neutrino masses. This is one of the main results of our paper.

After fixing the model parameters to fit the AMS-02 positron excess, we 
calculate the model predictions for total electron plus positron flux and find that it is consistent for the NH case with the recently released AMS-02 flux~\cite{AMS-update}, while the IH case is again disfavored as it gives a larger flux in the high energy part of 
the spectrum. The predictions for the diffuse photon flux in this 
model are also consistent with the latest Fermi-LAT results~\cite{fermi-gamma}. Finally, we justify the leptophilic nature of the DM candidate in our model, as 
required to explain the AMS-02 positron excess and the absence of a corresponding 
excess in the observed 
anti-proton flux from PAMELA~\cite{ap1, ap2}, which leads an upper limit on the DM-Higgs scalar quartic coupling, $\lambda_\Phi\lsim 0.06$. This implies an indirect upper limit on the DM-nucleon spin-independent scattering cross section of about $10^{-46}~{\rm cm}^2$ (for a DM mass of 1 TeV) which is roughly two orders of magnitude stronger than the current direct detection experimental bound from LUX~\cite{Akerib:2013tjd}.

The rest of the paper is organized as follows: In Section~\ref{sec2}, we briefly discuss the relevant features of the model and obtain a fit to the neutrino oscillation data for both normal and inverted hierarchies; in Section~\ref{sec2a}, we summarize the current experimental constraints in the triplet scalar sector, and in Section~\ref{sec2b}, we calculate the DM annihilation rate in this model. In Section~\ref{sec3}, we perform a $\chi^2$-fit to the AMS-02 positron fraction data for both NH and IH cases in two distinct kinematic regimes. In Section~\ref{sec4}, we compare our model predictions for various fluxes:  
total electron+positron flux, photon flux and neutrino flux with the corresponding experimental results. In Section~\ref{sec5}, we comment on the effect of a large DM-Higgs quartic coupling $\lambda_\Phi$ on our model predictions. Finally, our conclusions are given in Section~\ref{sec6}. 
\section{The Model}\label{sec2}
In the minimal type-II seesaw model, in addition to the SM fields, a triplet scalar field $\mathbf{\Delta}$ is introduced, which transforms as $({\bf 3},2)$ under the $SU(2)_L\times U(1)_Y$ gauge group: 
\begin{eqnarray}
\mathbf{\Delta} = \frac{\sigma_i}{\sqrt 2}\Delta_i = \left(\begin{array}{cc}
\Delta^+/\sqrt 2 & \Delta^{++}\\
\Delta^0 & -\Delta^+/\sqrt 2
\end{array}
\right),
\end{eqnarray}
where $\sigma_i$'s are the usual $2\times 2$ Pauli matrices, and  $\Delta_1=(\Delta^{++}+\Delta^0)/\sqrt 2,~\Delta_2=i(\Delta^{++}-\Delta^0)/\sqrt 2,~\Delta_3=\Delta^+$. The Yukawa Lagrangian for this model is given by 
\begin{eqnarray}
{\cal L}_Y &=& {\cal L}_Y^{\rm SM}-\frac{1}{\sqrt 2}\left(Y_\Delta\right)_{ij} L_i^{\sf T}Ci\sigma_2\mathbf{\Delta} L_j+{\rm H.c.},
\label{lag}
\end{eqnarray}
where $C$ is the Dirac charge conjugation matrix with respect to the Lorentz group, $(Y_\Delta)_{ij}$ denotes the Yukawa couplings in the lepton sector, and 
$L_i=(\nu_i,\ell_i)_L^{\sf T}$ (with $i=e,\mu,\tau$) is the $SU(2)_L$ lepton doublet. 

Following the notation used in~\cite{DGOS}, the scalar potential relevant 
for the minimal type-II seesaw can be written as
\begin{eqnarray} 
{\cal V}(\Phi,\Delta) &=& -m_\Phi^2(\Phi^\dag \Phi)+\frac{\lambda}{2}(\Phi^\dag \Phi)^2+M^2_\Delta {\rm Tr}(\mathbf{\Delta} ^\dag \mathbf{\Delta})
+ \frac{\lambda_1}{2}\left[{\rm Tr}(\mathbf{\Delta} ^\dag \mathbf{\Delta})\right]^2\nonumber\\
&& 
+\frac{\lambda_2}{2}\left(\left[{\rm Tr}(\mathbf{\Delta} ^\dag \mathbf{\Delta})\right]^2
-{\rm Tr}\left[(\mathbf{\Delta} ^\dag \mathbf{\Delta})^2\right]\right) +\lambda_4(\Phi^\dag \Phi){\rm Tr}(\mathbf{\Delta} ^\dag \mathbf{\Delta})+\lambda_5\Phi^\dag[\mathbf{\Delta}^\dag,\mathbf{\Delta}]\Phi\nonumber\\
&& 
+\left(\frac{\Lambda_6}{\sqrt 2}\Phi^{\sf T}i\sigma_2\mathbf{\Delta}^\dag \Phi+{\rm H.c.}\right)\, ,
\label{eq:Vpd}
\end{eqnarray}
where $\Phi=(\phi^+,\phi^0)^{\sf T}$ is the SM Higgs doublet. 
The coupling constants $\lambda_i$ can be chosen to be real through a phase redefinition of the field $\mathbf{\Delta}$. Also, we have chosen $m_\Phi^2>0$ in order to ensure the spontaneous symmetry breaking of the $SU(2)_L\times U(1)_Y$ gauge group to $U(1)_Q$ by a nonzero VEV of the neutral component of the SM Higgs doublet $\Phi$, 
$\langle \phi^0\rangle = v/\sqrt 2$ with $v\simeq 246.2$ GeV, while $M_\Delta^2$ can be of either sign.  

Note that the $\Lambda_6$ term in Eq.~(\ref{eq:Vpd}) is the {\it only} source of lepton number violation at the Lagrangian level before the spontaneous symmetry breaking.  A non-zero VEV for the Higgs doublet field $\Phi$ induces a tadpole term for the scalar triplet field $\mathbf{\Delta}$ via this term in Eq.~(\ref{eq:Vpd}), thereby generating a non-zero VEV for its neutral component, $\langle \delta^0\rangle = v_\Delta/\sqrt 2$, and breaking lepton number by two units. This results in the following Majorana mass matrix for the neutrinos: 
\begin{eqnarray}
(M_\nu)_{ij} = v_\Delta (Y_\Delta)_{ij}\, .
\label{eq:neutrino}
\end{eqnarray}
In order to satisfy the low-energy neutrino oscillation data~\cite{PDG}, 
we fix the structure of the Yukawa coupling matrix $Y_\Delta$ using Eq.~(\ref{eq:neutrino}):
\begin{eqnarray}
Y_\Delta = \frac{M_\nu}{v_\Delta} = \frac{1}{v_\Delta}U^{\sf T}M_\nu^{\rm diag}U \, ,
\label{YD}
\end{eqnarray}
where $M_\nu^{\rm diag} = {\rm diag}(m_1,m_2,m_3)$ is the diagonal neutrino mass eigenvalue matrix, and $U$ is the standard Pontecorvo-Maki-Nakagawa-Sakata (PMNS) mixing matrix which is usually parametrized in terms of the three mixing angles $\theta_{12},\theta_{23},\theta_{13}$, and one Dirac ($\delta$) and two Majorana ($\alpha_1,\alpha_2$) $C\!P$ phases:  
\begin{eqnarray}  
U = \left(\begin{array}{ccc}
c_{12}c_{13} & s_{12}c_{13} & s_{13}e^{-i\delta}\\
-s_{12}c_{23}-c_{12}s_{23}s_{13}e^{i\delta} &
c_{12}c_{23}-s_{12}s_{23}s_{13}e^{i\delta} & s_{23}c_{13}\\ 
s_{12}s_{23}-c_{12}c_{23}s_{13}e^{i\delta} &
-c_{12}s_{23}-s_{12}c_{23}s_{13}e^{i\delta} & c_{23}c_{13} 
\end{array}\right)\times{\rm
  diag}(e^{i\alpha_1/2},e^{i\alpha_2/2},1)\; \nonumber,
\end{eqnarray}
where $c_{ij}\equiv \cos\theta_{ij},~s_{ij}\equiv \sin\theta_{ij}$. 
For our illustration purposes, we assume the Majorana phases in the PMNS matrix to be zero, and use the 
central values of a recent global analysis of the 3-neutrino oscillation data~\cite{global}:
\begin{eqnarray}
&&
\Delta m^2_{\rm sol}=  7.62\times 10^{-5}~{\rm eV}^2,~
\Delta m^2_{\rm atm} = 2.55\times 10^{-3}~{\rm eV}^2,~\nonumber\\
&&
\theta_{12}=34.4^\circ,~ \theta_{23}=40.8^\circ,\theta_{13}=9.0^\circ,
\delta=0.8\pi
\end{eqnarray} 
to obtain the following structure of the Yukawa coupling matrix from Eq.~(\ref{YD}): 
\begin{eqnarray}
Y_\Delta &=& \frac{10^{-2}~{\rm eV}}{v_\Delta}\times \left\{\begin{array}{cc}
\left(\begin{array}{ccc}
0.31-0.12i & -0.09+0.32i & -0.72+0.37i \\
-0.09+0.32i & 2.53+0.04i & 2.19+0.01i \\
-0.72+0.37i & 2.19+0.01i & 3.07-0.03i
\end{array}\right) & ({\rm normal~hierarchy}) \\
\left(\begin{array}{ccc}
4.95+0i &  0.44 + 0.30 i & 0.46 + 0.35 i \\
0.44 + 0.30 i &  2.94 + 0.05 i & -2.50 + 0.06 i \\
0.46 + 0.35 i & -2.50 +  0.06 i & 
2.20 + 0.06 i \end{array}\right) & ({\rm inverted~hierarchy})
\end{array}\right.
\label{yuk}
\end{eqnarray} 
Here we have chosen the lightest neutrino mass to be zero so that the other two mass eigenvalues in Eq.~(\ref{YD}) are completely fixed by the solar and atmospheric mass-squared differences. However, our final results will not depend on the absolute value of the lightest neutrino mass, unless it is close to the degenerate limit, $m_{\rm i} \gsim 0.1$ eV, in which case, we have democratic Yukawa couplings of the triplet scalar to all the charged-leptons. We do not consider this special case in our analysis, since this is in conflict with the most stringent upper limit on the sum of light neutrino masses from Planck data: $\sum_i m_i < 0.23$ eV at 95\% CL~\cite{Planck}.  

For the scalar sector of the minimal type-II seesaw model, after the neutral fields $\phi^0$ and $\Delta^0$ acquire VEVs, we have seven physical mass 
eigenstates $H^{\pm\pm},H^\pm,h,H^0,A^0$ with the following eigenvalues:
\begin{eqnarray}
m^2_{H^{\pm\pm}} &=& M_\Delta^2+\frac{1}{2}(\lambda_4+\lambda_5)v^2+\frac{1}{2}(\lambda_1+\lambda_2)v_\Delta^2, \label{doub-mass}\\
m^2_{H^\pm} &=& \left(M_\Delta^2+\frac{1}{2}\lambda_4v^2+\frac{1}{2}\lambda_1v_\Delta^2\right)\left(1+\frac{2v^2_\Delta}{v^2}\right),\label{sing-mass}\\
m^2_{A^0} &=& \left(M_\Delta^2+\frac{1}{2}(\lambda_4-\lambda_5)v^2+\frac{1}{2}\lambda_1v_\Delta^2\right)\left(1+\frac{4v^2_\Delta}{v^2}\right), \label{A-mass}\\
m^2_{H^0} &=& \frac{1}{2}\left(A+C+\sqrt{(A-C)^2+4B^2}\right), \label{Hz-mass}\\
m^2_h &=& \frac{1}{2}\left(A+C-\sqrt{(A-C)^2+4B^2}\right), \label{h-mass}\\
{\rm with}~~
A = \lambda v^2,&&
B = -\frac{2v_\Delta}{v}\left(M^2_\Delta+\frac{1}{2}\lambda_1 v_\Delta^2\right),~~
C = M^2_\Delta+\frac{1}{2}(\lambda_4-\lambda_5)v^2+\frac{3}{2}\lambda_1 v_\Delta^2.\nonumber
\label{eq:scamass}
\end{eqnarray}
In the limit $v_\Delta\ll v$, the mixing between the doublet and triplet scalars is usually small (unless the $C\!P$-even scalars $h$ and $H^0$ are close to being mass-degenerate). In this limit, the mass of the (dominantly doublet) lightest $C\!P$-even scalar is simply given by $m_h^2=\lambda v^2$ (as in the SM) independent of the mass scale $M_\Delta$, whereas the other (dominantly triplet) scalars have $M_\Delta$-dependent mass. 
\subsection{Experimental Constraints}\label{sec2a}
The triplet VEV contributes to the SM $W$- and $Z$-boson masses at the tree-level thereby affecting the $\rho$-parameter. The constraints on the $\rho$-parameter from the electroweak precision data~\cite{PDG} require the triplet VEV to be very small compared to the electroweak VEV: $v_\Delta< 2$ GeV~\cite{delAguila:2008ks}. 
On the other hand, the singly- and doubly-charged scalars in the type-II seesaw model contribute to the 
lepton flavor violating (LFV) processes such as $\ell^-_i\to \ell^-_j\gamma$ and $\ell^-_i\to \ell^-_j\ell^+_k\ell^-_l$, and also to the muon anomalous magnetic moment. 
The current experimental limits on LFV put a stringent lower bound~\cite{lfv1, lfv2} 
\begin{eqnarray}
 v_\Delta M_\Delta\gsim 150~{\rm eV}~{\rm GeV}.
\label{lfv}
\end{eqnarray} 

There also exist strong lower limits on the type-II seesaw scale from collider searches for doubly-charged Higgs bosons decaying to pairs of prompt, isolated, high-$p_T$ electrons/muons with the same electric charge~\cite{ATLAS, CMS}. The masses of doubly-charged Higgs bosons are constrained depending on the branching ratio into these leptonic final states. Assuming 100\% branching ratio for leptonic final states, which is true for 
large Yukawas or small $v_\Delta\lsim 0.1$ MeV, the current lower limits are in the range of 375 - 409 GeV depending on the flavor of the final-state lepton-pair~\cite{ATLAS}. These limits are significantly weakened for $v_\Delta> 0.1$ MeV as the leptonic branching ratio becomes small, and could be as low as 100 GeV in certain cases~\cite{goran}.
\subsection{Dark Matter}\label{sec2b}
The minimal type-II seesaw model can be extended to accommodate a cold DM candidate by simply adding a SM singlet real scalar field $D$~\cite{Gogoladze:2009gi}. Its stability can be ensured by assigning it an odd $Z_2$-parity, whereas the SM lepton and Higgs doublets as well as the complex scalar triplet are assigned an even $Z_2$-parity. This is summarized in Table~\ref{tab1}. The scalar potential relevant for the DM physics is given by\begin{eqnarray}
{\cal V}_{\rm DM}(\Phi,\mathbf{\Delta}, D) &=& \frac{1}{2}m_D^2D^2+\lambda_D D^4+\lambda_\Phi D^2(\Phi^\dag \Phi)+\lambda_\Delta D^2{\rm Tr}(\mathbf{\Delta}^\dag \mathbf{\Delta})
\label{dm}
\end{eqnarray}
which can be rewritten in terms of the physical scalar mass eigenstates, as 
follows:
\begin{eqnarray}
{\cal V}_{\rm DM} &=& \frac{1}{2}m_{\rm DM}^2D^2+\lambda_D D^4+\lambda_\Phi vD^2h+\frac{1}{2}\lambda_\Phi D^2h^2
\nonumber\\
&& +\lambda_\Delta D^2\left(H^{++}H^{--}+H^+H^-+\frac{1}{2}\left[(H^0)^2+(A^0)^2\right]+v_\Delta H^0\right)
\label{dm2}
\end{eqnarray}
assuming negligible mixing between the doublet and triplet scalars. Here $m_{\rm DM}^2=m_D^2+\lambda_\Phi v^2+\lambda_\Delta v_\Delta^2$ is the physical mass of the DM candidate in this model.  
\begin{table}[ht]
\begin{center}
\begin{tabular}{c|c|c|c}\hline\hline
Field & $SU(2)_L$ & $U(1)_Y$ & $Z_2$ \\ \hline
$\mathbf{\Delta}$  & {\bf 3} & 2 & + \\
$\Phi$ & {\bf 2} & 1 & +\\ 
$\nu_{\ell_L}$ & {\bf 2} & 1 & +\\
$\ell_L$ & {\bf 2} & $-1$ & +\\
$D$ & {\bf 1} & 0 & $-$\\ 
\hline\hline
\end{tabular}
\end{center}
\caption{The relevant particle content of the minimal Type-II seesaw+DM model.}
\label{tab1}
\end{table}

Thus for $m_{\rm DM}$ larger than the doublet and triplet masses, the DM candidate in this model can {\it directly} annihilate into a pair of scalars through the quartic couplings $\lambda_\Delta$ and $\lambda_\Phi$ in Eq.~(\ref{dm2}). In the non-relativistic (cold DM) limit $v^2\ll 1$, the annihilation rate is given by 
\begin{eqnarray}
\langle \sigma v\rangle = \frac{1}{16\pi m_{\rm DM}^2}\left[\lambda_\Phi^2\sqrt{1-\frac{m_h^2}{m_{\rm DM}^2}}+6\lambda_\Delta^2\sqrt{1-\frac{m_\Delta^2}{m_{\rm DM}^2}}\right]
\label{an}
\end{eqnarray}
which should match the thermal annihilation rate $\langle \sigma v\rangle_{\rm th}\simeq 1$ pb in order to reproduce the observed cold DM relic density of $\Omega_{\rm CDM}h^2=0.1199\pm 0.0027$~\cite{Planck}.\footnote{This correlation between the annihilation rate $\langle \sigma v\rangle $ and relic density $\Omega h^2$ is only valid for a thermal DM candidate. If the DM has a non-thermal evolution history, its final relic abundance can be completely set by initial conditions, irrespective of the value of $\langle \sigma v\rangle $ (see e.g.,~\cite{Gelmini:2006pw, Easther:2013nga, Dev:2013yza}).} Note that $m_\Delta$ in Eq.~(\ref{an}) generically represents the triplet scalar masses given by 
Eqs.~(\ref{doub-mass}-\ref{Hz-mass}) which we have assumed here to be degenerate, and hence, the factor of 6 in the second term of Eq.~(\ref{an}).  
 
For $\lambda_\Delta\gg \lambda_\Phi$, the DM-pair dominantly annihilates to scalar 
triplets: 
\begin{eqnarray}
DD\to H^{++}H^{--},H^{+}H^{-},H^0H^0,A^0A^0.
\end{eqnarray} 
These scalar triplets 
subsequently decay to leptonic final states: $H^{++}\to {\ell^+\ell^+}$,  
$H^+\to \ell^+ \nu_\ell$, and $H^0, A^0\to \nu\nu$  with almost 100\% branching ratio provided $v_\Delta\lsim 0.1$ MeV. The final-state lepton-flavor depends on the structure of the Yukawa coupling matrix in Eq.~(\ref{yuk}). Thus 
the DM candidate in our model is `leptophilic' for a wide range of model parameter space, 
and hence, can give rise to a positron excess without a corresponding anti-proton excess. 
   
On the other hand, for a large quartic coupling $\lambda_\Phi$, the DM-pair 
annihilates to SM Higgs bosons: $DD\to hh$ and each Higgs boson subsequently  
decays mostly to bottom quarks and gauge bosons (with branching ratios of about 
58\% and 22\% respectively for $m_h=125$ GeV). This would give an excess of 
anti-proton flux over cosmic-ray background which is incompatible with the observed anti-proton flux from PAMELA~\cite{ap2}. Moreover, a large value of $\lambda_\Phi$ would also lead to a large direct 
detection cross section through  elastic scattering of $D$ off 
nuclei mediated by a SM Higgs boson in the $t$-channel, and would be in conflict with the observed lower limits from LUX~\cite{Akerib:2013tjd}.  For these reasons, we assume $\lambda_\Phi\ll \lambda_\Delta$  in Eq.~(\ref{an}) for our 
subsequent analysis. We will comment on the implications of a large 
$\lambda_\Phi$ on our results for positron fraction fit and the 
anti-proton flux in Section~\ref{sec5}. 

It is worth mentioning here that the model we have considered here is the {\it minimal}  extension of the Type-II seesaw to accommodate a DM candidate, while maintaining a connection with the neutrino mass spectrum. One can also consider a model with fermionic DM $\chi$, instead of a scalar DM in Eq.~(\ref{dm}), to explain the AMS-02 positron data within this setup. However, in the fermionic DM case, one needs to have an additional field, e.g. a scalar singlet $S$, which couples to $\chi$ as well as to the triplets, or a dimension-5 operator with the direct $\bar{\chi}\chi {\mathbf{\Delta}}^\dag{\mathbf{\Delta}}$ coupling, in order to allow for the DM annihilation to the triplets, whose further decay to leptons carry the information on the Yukawa couplings. Apart from this basic difference, our subsequent analysis should also apply to the fermionic DM case.
\section{Fitting the AMS-02 data}\label{sec3}
In our model, the DM pair-annihilation into a pair of doubly (singly)-charged $SU(2)_L$-triplet scalars and their subsequent decay produces leptonic final states with 
four (two) charged leptons. The flavor of the final-state lepton is determined by the structure of the Yukawa coupling as given in Eq.~(\ref{yuk}). For a 
normal hierarchy of neutrino masses, the triplets will decay dominantly to  
muon and tau final states, whereas for inverted hierarchy, they more often decay to an electron final state.  
If the triplets decay mainly to electron/positron final states, the positron flux due to the DM 
annihilation will sharply rise above the cosmic-ray background as we go to higher energies, and will eventually drop to the background level before $E=m_{\rm DM}$. On the other hand, if they mostly decay to muons and taus (which subsequently decay to electrons with their SM branching ratio), the final positron flux will have a much softer 
energy spectrum, before eventually dropping to the background level for some $E<m_{\rm DM}$. Thus for a given DM mass, we would expect the positron fraction for a normal hierarchy of neutrino masses to be softer towards the higher energy end as compared to 
that for an inverted hierarchy in our model. Since the AMS-02 positron data~\cite{AMS} indeed 
exhibits a softer energy spectrum in the high energy regime,\footnote{The slope of the AMS-02 positron fraction spectrum decreases by an order of magnitude from 20 GeV to 250 GeV~\cite{AMS-update}.} as compared to PAMELA~\cite{pamela1, pamela2, pamela-update} and Fermi-LAT~\cite{fermi} results, the fit to the AMS-02 data is expected to be much better for a normal hierarchy of 
neutrinos, as we show later explicitly.       

Before we discuss our fit results, it is important to mention here that we still need a DM 
annihilation cross section much larger than the thermal cross section of $\langle \sigma v\rangle_{\rm th}\simeq  1$ pb, in order to fit the AMS-02 data. This difference can be reconciled by the so-called `boost-factor' which could have an astrophysical origin due to small-scale inhomogeneities of the DM density distribution in the galactic halo. Note that even the highest resolution numerical simulations can only resolve a fraction of this small-scale substructure, and it depends sensitively on subhalo properties orders of magnitude below the current resolution limit of numerical simulations 
(for a review of the current state of the art, see Ref.~\cite{Kuhlen:2012ft}). 
Since the DM annihilation rates are proportional to the square of its density, 
any unresolved small-scale clumpiness will significantly enhance the annihilation rate. Alternatively, one could also have a particle physics origin of the boost factor 
due to the Breit-Wigner enhancement of DM annihilation~\cite{bw1, bw2, bw3}. In this mechanism, the DM particles in the present Universe pair-annihilate efficiently through an $s$-channel process mediated by a particle with mass very close to but slightly smaller than twice the DM mass, while the same process can be relatively suppressed in the early Universe due to a large relative velocity between the DM particles at freeze-out which pushes their total energy away from the $s$-channel resonance pole. This mechanism can easily be implemented in our 
model by introducing another $Z_2$-even singlet real scalar field $S$ coupling to both $D$ and $\mathbf{\Delta}$ such that the DM annihilation can proceed via an $s$-channel resonance: $DD\to S\to \mathbf{\Delta}^\dag \mathbf{\Delta}$ for a suitable range of masses for $S$~\cite{Gogoladze:2009gi}. For our purposes, we can just parametrize this enhancement by a boost factor $B=\langle \sigma v\rangle /\langle\sigma v\rangle_{\rm th}$ such that we reproduce the thermal annihilation cross 
section, and hence, the observed DM relic density, irrespective of the actual origin of the boost factor.  
\subsection{Positron Fraction} \label{sec:3a}
We implement our model Lagrangian in {\tt CalcHEP3.4}~\cite{Belyaev:2012qa}, and calculate the positron flux due to DM annihilation $\left(\Phi_{e^+}^{\rm DM}\right)$ using {\tt micrOMEGAs3.2}~\cite{Belanger:2013oya}. For the DM density distribution in the galactic halo, we have used the Navarro-Frenk-White (NFW) density profile~\cite{nfw1, nfw2}. The galactic propagation of charged particles and the modulation effects due to solar wind are taken into account using the default global parameters of the indirect detection module in {\tt micrOMEGAs3.2}~\cite{Belanger:2013oya, Belanger:2010gh}. The excess positron fraction is then computed by comparing this positron flux with 
the known cosmic-ray primary electron and secondary electron plus positron background. These background fluxes are obtained by solving the diffusion equation for cosmic-ray particles using a widely used galactic cosmic-ray propagation model due to Moskalenko and Strong~\cite{Moskalenko:1997gh, strong} which assumes a homogeneous source distribution with cylindrical symmetry. To take into account the electron energy losses by ionization, Coulomb interactions, bremsstrahlung, inverse Compton scattering and synchrotron radiation, we use {\tt GALPROPv54}~\cite{galprop} for which the most 
relevant input parameters used in our analysis were chosen as follows~\cite{fermi-flux2}: an electron injection index of 2.5 for $E> 4$ GeV and 1.6 for $E\leq 4$ GeV with a modulation potential of 550 MV, a spatial Kolmogorov diffusion with diffusion coefficient of $5.75\times 10^{28}~{\rm cm}^2{\rm s}^{-1}$ and spectral index of 0.33, a diffusive reacceleration characterized by an Alfv\'{e}n speed of 30 km s$^{-1}$, and a halo radius of 8.5 kpc. The background fluxes obtained for these {\tt GALPROP} input parameters are roughly consistent with the following parametrization~\cite{Baltz:1998xv} in the energy 
range 10-1000 GeV:
\begin{eqnarray}
	\Phi_{e^-}^{(\rm prim)}(E) &=& \frac{0.16E^{-1.1}}{1+11E^{0.9}+3.2E^{2.15}}{\rm GeV}^{-1}{\rm cm}^{-2}{\rm s}^{-1}{\rm sr}^{-1}, \label{pe}\\
	\Phi_{e^-}^{(\rm sec)}(E) &=& \frac{0.70E^{0.7}}{1+110E^{1.5}+600E^{2.9}+580E^{4.2}}{\rm GeV}^{-1}{\rm cm}^{-2}{\rm s}^{-1}{\rm sr}^{-1}, \label{se}\\
	\Phi_{e^+}^{(\rm sec)}(E) &=& \frac{4.5E^{0.7}}{1+650E^{2.3}+1500E^{4.2}}{\rm GeV}^{-1}{\rm cm}^{-2}{\rm s}^{-1}{\rm sr}^{-1},\label{sp} 
\end{eqnarray}   
(where $E$ is in units of GeV). The total electron plus positron flux and the 
positron fraction in our model are respectively given by 
\begin{eqnarray}
 \Phi_{\rm tot} &\equiv& \Phi^{\rm bkg}+\Phi^{\rm DM} = \kappa \left(\Phi_{e^-}^{(\rm prim)}+\Phi_{e^-}^{(\rm sec)}+\Phi_{e^+}^{(\rm sec)}\right)+\left(\Phi_{e^-}^{\rm DM}+\Phi_{e^+}^{\rm DM}\right), \label{phitot}\\
f_{e^+}&\equiv& \frac{\Phi_{e^+}}{\Phi_{\rm tot}} =\frac{\Phi_{e^+}^{\rm DM}+\kappa \Phi_{e^+}^{(\rm sec)}}{\kappa \left(\Phi_{e^-}^{(\rm prim)}+\Phi_{e^-}^{(\rm sec)}+\Phi_{e^+}^{(\rm sec)}\right)+\left(\Phi_{e^-}^{\rm DM}+\Phi_{e^+}^{\rm DM}\right)},
\label{frac}
\end{eqnarray} 
where the scaling factor $\kappa$ is a free parameter which takes into account the uncertainty in the normalization of the astrophysical background. 

For a given DM mass ($m_{\rm DM}$) and triplet scalar masses ($m_{H^\pm},m_{H^{\pm\pm}}$), the scalar 
coupling $\lambda_\Delta$ can be fixed using Eq.~(\ref{an}) (and assuming $\lambda^2_\Phi\ll \lambda^2_\Delta$) to reproduce the observed DM thermal relic abundance. 
For instance, for $m_{\rm DM}=1$ TeV and 
$m^2_{\rm DM}\gg m^2_\Delta$, we require $\lambda_\Delta=0.15$ to obtain $\langle \sigma v\rangle=3\times 10^{-26}~{\rm cm}^3{\rm s}^{-1}$. Then we vary the free parameter $\kappa$ in Eq.~(\ref{frac}) 
and perform a $\chi^2$-minimization taking into account the AMS-02 positron fraction data~\cite{AMS} to determine the best-fit value of the boost factor for the corresponding model parameters. Note that the positron fraction $f_{e^+}$ in Eq.~(\ref{frac}) is independent of $\kappa$ in the sense that for a different value of $\kappa$, the boost factor (and hence $\Phi^{\rm DM}_{e^+}$) can be chosen accordingly to give the same $f_{e^+}$ without changing the goodness of the fit for a given set of model parameters. However, the choice of $\kappa$ will affect the model predictions for the total positron 
flux ($\Phi^{\rm DM}_{e^+}+\Phi^{\rm (sec)}_{e^+}$) and the total electron+positron flux ($\Phi_{\rm tot}$) which are also required to be consistent with the updated AMS-02 results~\cite{AMS-update}. We find that this requirement in our 
model favors values of $\kappa$ in the range 0.8 - 0.9 with the best-fit value around $\kappa=0.85$ which we will use in our subsequent analysis unless otherwise specified.       

For our illustration purposes, we choose two kinematically distinct scenarios: 
(a) non-degenerate 
case ($m^2_{\rm DM}\gg m^2_\Delta$)  where the triplet scalars produced by DM annihilation are highly boosted and the final-state leptons produced from $\mathbf{\Delta}$ decay have a continuous energy spectrum; (b) degenerate case ($m_{\rm DM}\simeq m_\Delta$) where the triplets produced by DM annihilation are almost at rest, and hence, the leptons coming from their decay have a monochromatic energy spectrum. For numerical purposes in case (a), we 
have chosen a common value of 450 GeV for the doubly- and singly-charged scalars, keeping in mind the current lower limits on the doubly-charged scalars which are in the range of 375 - 409 GeV~\cite{ATLAS} depending on the final-state lepton flavor. For case (b), we choose $\Delta m \equiv m_{\rm DM}-m_\Delta=10$ GeV (smaller values of $\Delta m$ could lead to  computational complications in {\tt CalcHEP}). For normal and inverted hierarchies of neutrino masses, we use the corresponding Yukawa couplings given by Eq.~(\ref{yuk}) to satisfy the neutrino oscillation data. For numerical purposes, we have chosen $v_\Delta=1$ eV so that the LFV constraints given by Eq.~(\ref{lfv}) can be satisfied even for a 
low seesaw scale; however, our results are independent of the exact value of $v_\Delta$ as long as $v_\Delta\lsim 0.1$ MeV so that the leptonic branching ratio for the $\mathbf{\Delta}$'s is almost 100\%. 

\begin{table}[h!]
\begin{center}
\begin{tabular}{c|c|c|c|c}\hline\hline
Case & $m_{\rm DM}$ (TeV) & boost factor & $\chi^2_{\rm min}$ & $\chi^2_{\rm min}$/dof \\ \hline
& 0.8 & 3320.23  & 78.41 & 2.31\\
& 0.9 & 4031.68 & 52.43 & 1.54 \\
& 1.0 & 4801.81 & 36.87 & 1.08\\
NH Case (a) & 1.1 & 5632.09 & 27.79 & 0.82 \\
& 1.2 & 6513.54 & 22.79 & 0.67 \\
& 1.3 & 7452.6 & 20.87 & 0.61 \\
& 1.4 & 8440.56 & 20.76 & 0.61 \\ 
& 1.5 & 9488.91 & 22.36 & 0.66 \\
\hline 
& 0.8 &3134.62  & 86.99 & 2.56\\
& 0.9 & 3774.12 & 55.68 & 1.64 \\
& 1.0 & 4467.27 &37.95 & 1.12\\
NH Case (b) & 1.1 & 5214.3 & 27.91 & 0.82 \\
& 1.2 & 6023.57 & 22.45 & 0.66\\
& 1.3 & 6870.72 & 21.58 & 0.63\\
& 1.4 & 7769.48 & 23.07 & 0.68 \\ 
& 1.5 & 8724.95 & 25.88 & 0.76 \\
\hline
& 0.5 & 976.37 &  114.49 & 3.37 \\
& 0.6 & 1368.43 & 113.57 & 3.34 \\
& 0.7 & 1823.37 & 123.48 & 3.63 \\
IH Case (a) & 0.8 &2481.02 & 135.82 & 3.99\\
& 0.9 & 2915.16 & 148.15 & 4.36 \\
& 1.0 & 3755.20 & 159.64 & 4.69\\ \hline 
& 0.5 & 1031.94 & 129.33 & 3.50\\
& 0.6 &1374.9 & 119.42 & 3.51 \\
& 0.7 & 1817.25 & 141.08 & 4.15 \\
IH Case (b) & 0.8 &2354.59 &155.81 & 4.58\\
& 0.9 & 2842.86 & 175.11 & 5.15 \\
& 1.0 &3365.19 & 192.76 & 5.67\\ \hline 
\hline
\end{tabular}
\end{center}
\caption{The best-fit values of the boost factor along with the corresponding minimum $\chi^2$ and $\chi^2$ per degrees of freedom (dof=34) for both NH and IH cases, and in two kinematic regimes: Case (a) $m^2_\Delta\ll m^2_{\rm DM}$ and (b) $m_{\rm DM}-m_\Delta=10$ GeV. Here we have chosen $\kappa=0.85$ in Eq.~(\ref{frac}). }
\label{tab2}
\end{table}    
The best-fit values of the boost factor from our $\chi^2$-minimization are given in Table~\ref{tab2} for different DM masses. The $\chi^2$-value is defined as 
\begin{eqnarray}
\chi^2=\sum_i\frac{\left(f_i^{\rm model}-f_i^{\rm AMS}\right)^2}{\delta f_i^2}
\end{eqnarray}
where $f_i$'s are the relevant observables (positron fraction in this case), $\delta f_i$'s are the corresponding experimental errors (stat+syst) taken from~\cite{AMS}, and $i$ runs over all the available data points. We have used the AMS-02 data points for $E>15$ GeV (36 in total) in our $\chi^2$-analysis. Including the low-energy data points (below 15 GeV) gives a poor fit, but we note that 
the discrepancies in the low-energy region could be accounted for by uncertainties caused by solar modulation~\cite{Asaoka:2001fv}, as well as due to background flux uncertainties. We have not used any other previous data sets for positron fraction from Fermi-LAT, PAMELA etc, since the precision of the AMS-02 data set is much better. We also neglect the finite energy resolution effect of the electromagnetic calorimeter since it is very small for the AMS-02 detector~\cite{AMS} (less than 3\% for $E>15$ GeV). It is clear from Table~\ref{tab2} that the $\chi^2$-values, and hence, the goodness of fit for NH are much better than those for IH in both cases (a) and (b). Also for NH, higher DM masses are preferred with the $\chi^2$-value decreasing at higher masses, 
whereas for IH, lower DM masses are preferred with the $\chi^2$-value increasing with DM mass. In both cases, the boost factor increases with increase in the DM mass in order to be able to fit the AMS-02 data in the low-energy bins which have much smaller error bars.    
\begin{figure}[t]
\centering
\includegraphics[width=6.5cm]{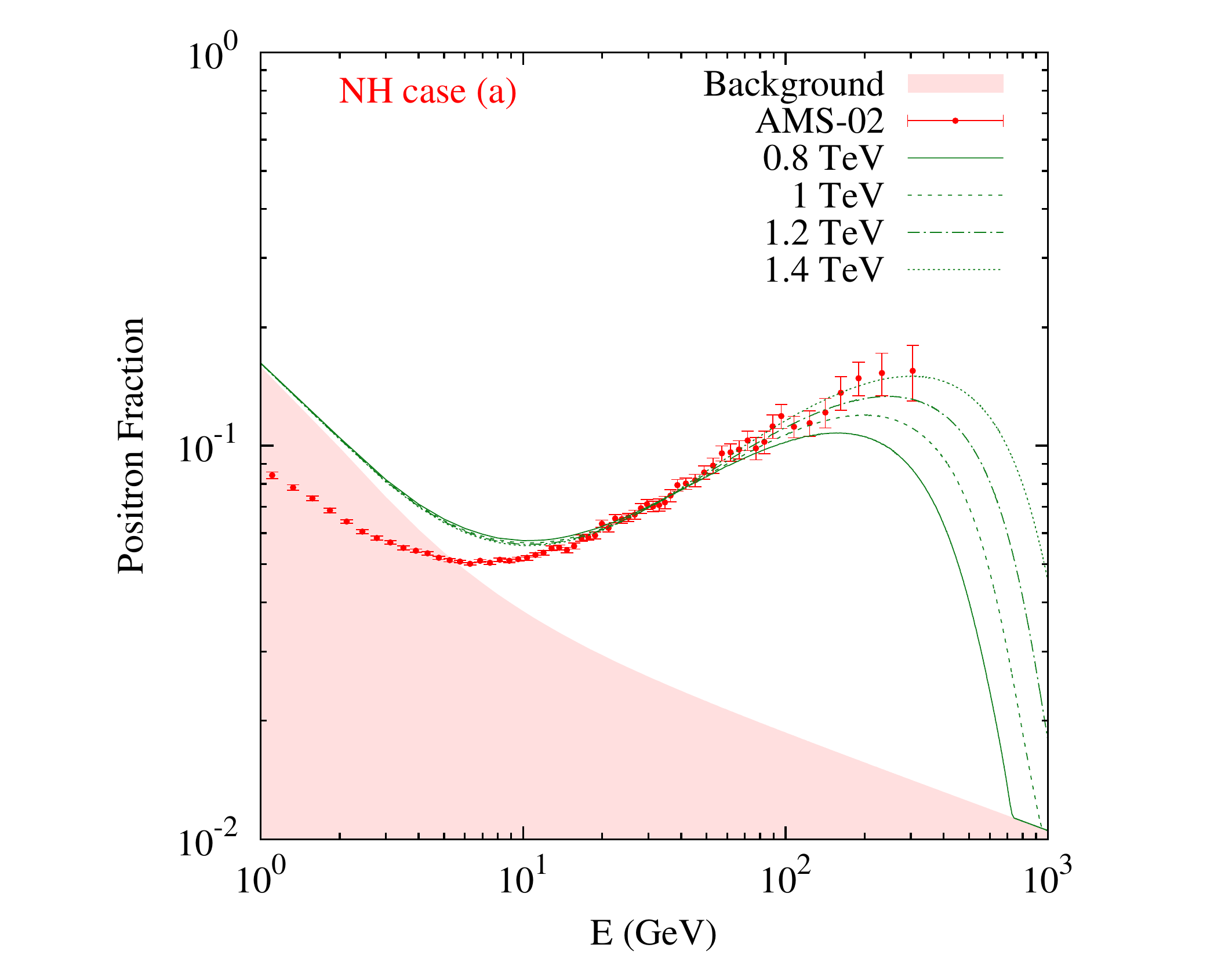}
\hspace{0.5cm}
\includegraphics[width=6.5cm]{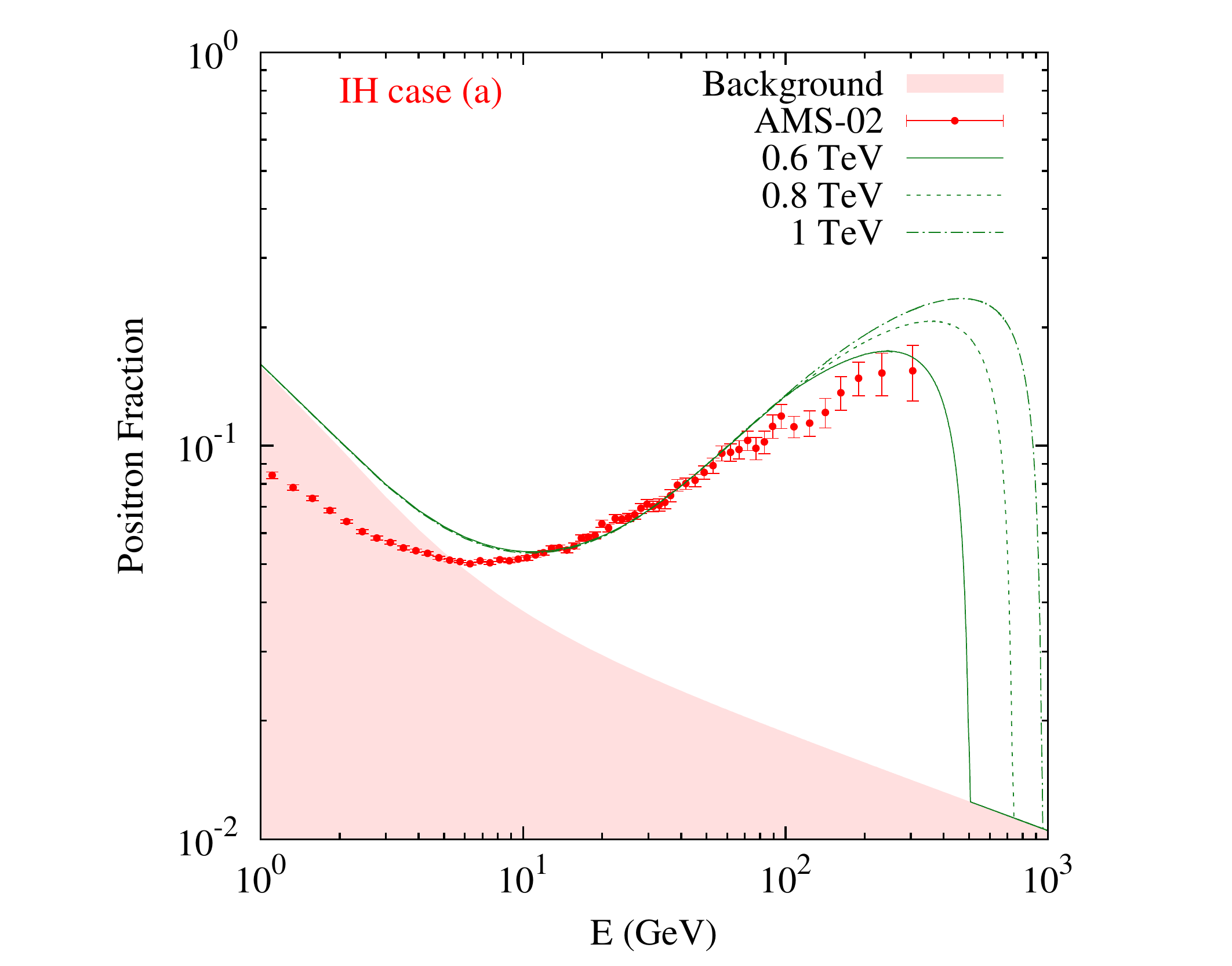}\\
\includegraphics[width=6.5cm]{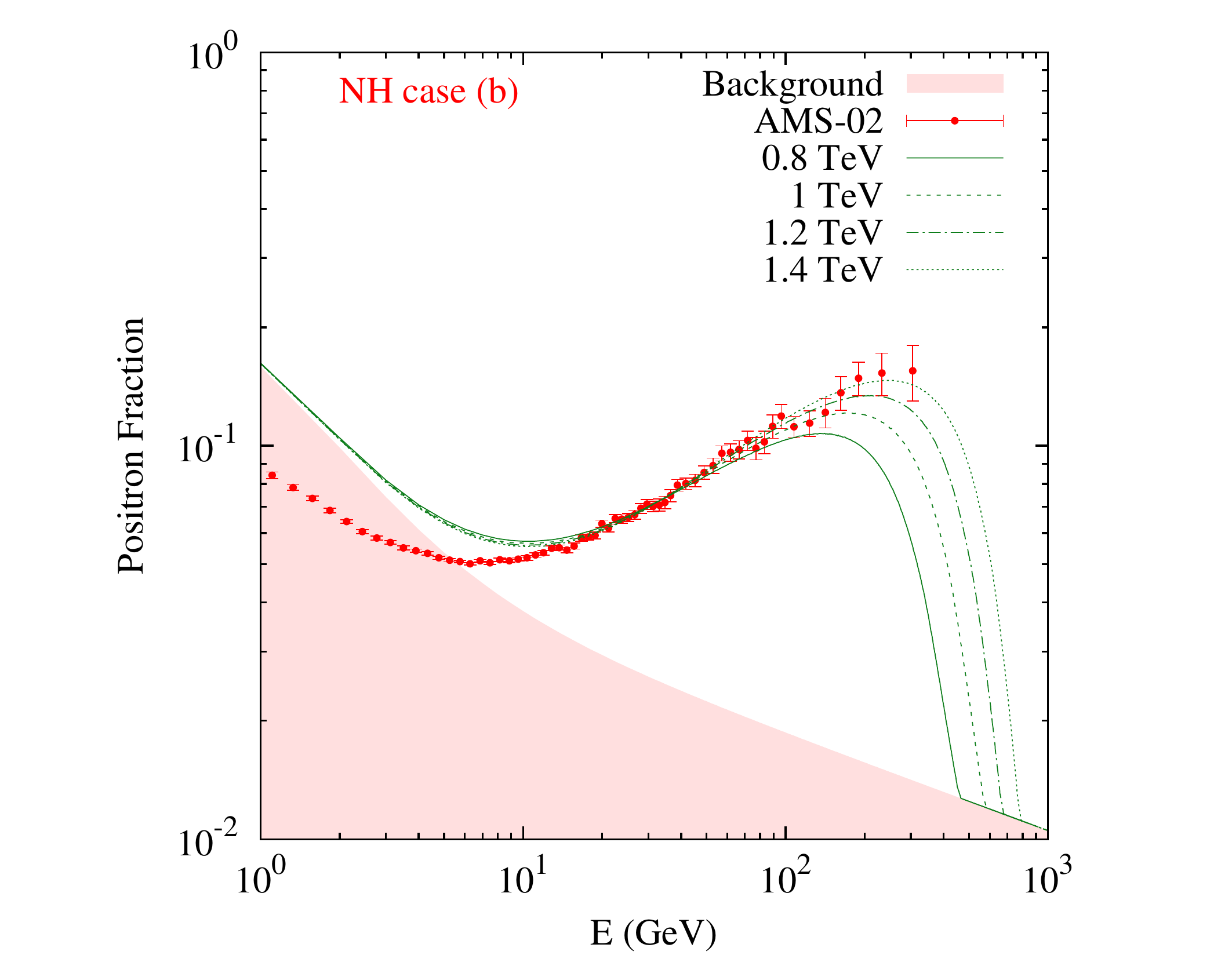}
\hspace{0.5cm}
\includegraphics[width=6.5cm]{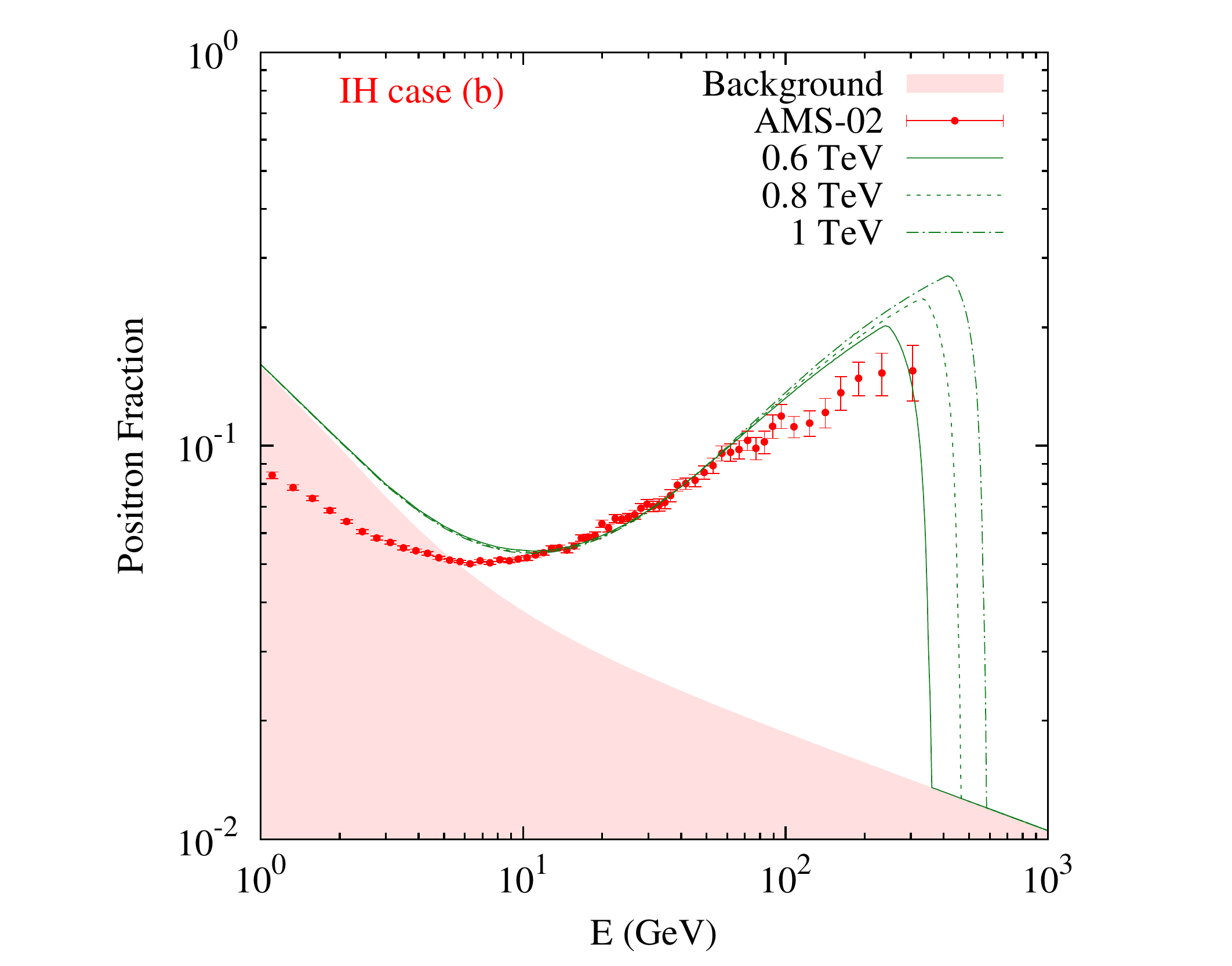}
\caption{The positron fraction in our model as a function of the positron energy for various DM masses (in TeV). The 
left (right) panels are for NH (IH) of neutrino masses, whereas the top and bottom panels show the results for 
case (a) with $m^2_\Delta\ll m^2_{\rm DM}$ and for case (b) with $m_{\rm DM}-m_\Delta=10$ GeV respectively. 
Also shown are the AMS-02 data~\cite{AMS} (with error bars) and the secondary cosmic-ray positron background (shaded region) for comparison.}
\label{fig1}
\end{figure}

We compare our fit results for the positron fraction with the AMS-02 data in Fig.~\ref{fig1} for both NH (left panels) and IH (right panels) and for both 
cases (a) and (b). We also show the background positron fraction due to secondary cosmic-ray positron flux as in Eq.~(\ref{sp}). Thus we find that the observed positron excess over expected cosmic-ray background can be explained in our model for $E>15$ GeV for both cases (a) and (b). The discrepancy between the AMS-02 data and our fit values below 15 GeV could be accounted for by solar modulation effects which are at the level of a few percent at 10 GeV, increasing to about 30\% at 1 GeV~\cite{DM1}. From Table~\ref{tab2} as well as from Fig.~\ref{fig1}, it is clear that the NH fits are much better compared to the IH fits.  This is due to the fact that the positron energy spectrum is harder for the IH case since it comes directly from the $\mathbf{\Delta}$ decay, whereas in the NH case, the spectrum is softer as it comes from muon and tau decays. 
Since the rise in the AMS-02 positron fraction becomes softer toward higher energies, it is more difficult to fit both low- and high-energy bins for the IH case, as compared to the NH case. Also note that for case (b) with almost degenerate DM and triplet masses, the positron energy spectrum rises sharply with a sudden cut-off at high energy due to the fact that they are produced from $\mathbf{\Delta}$'s which are almost at rest. This feature is more prominent for the IH case again due to the fact that the positrons are directly produced from the $\mathbf{\Delta}$ decay. Note that the fall-off behavior of the positron fraction at some higher energy is a generic feature of the single-component DM interpretation, and can be tested with more data at higher energies.
\subsection{Allowed Parameter Space}\label{sec:3b}
\begin{figure}[t]
\centering
\includegraphics[width=6.5cm]{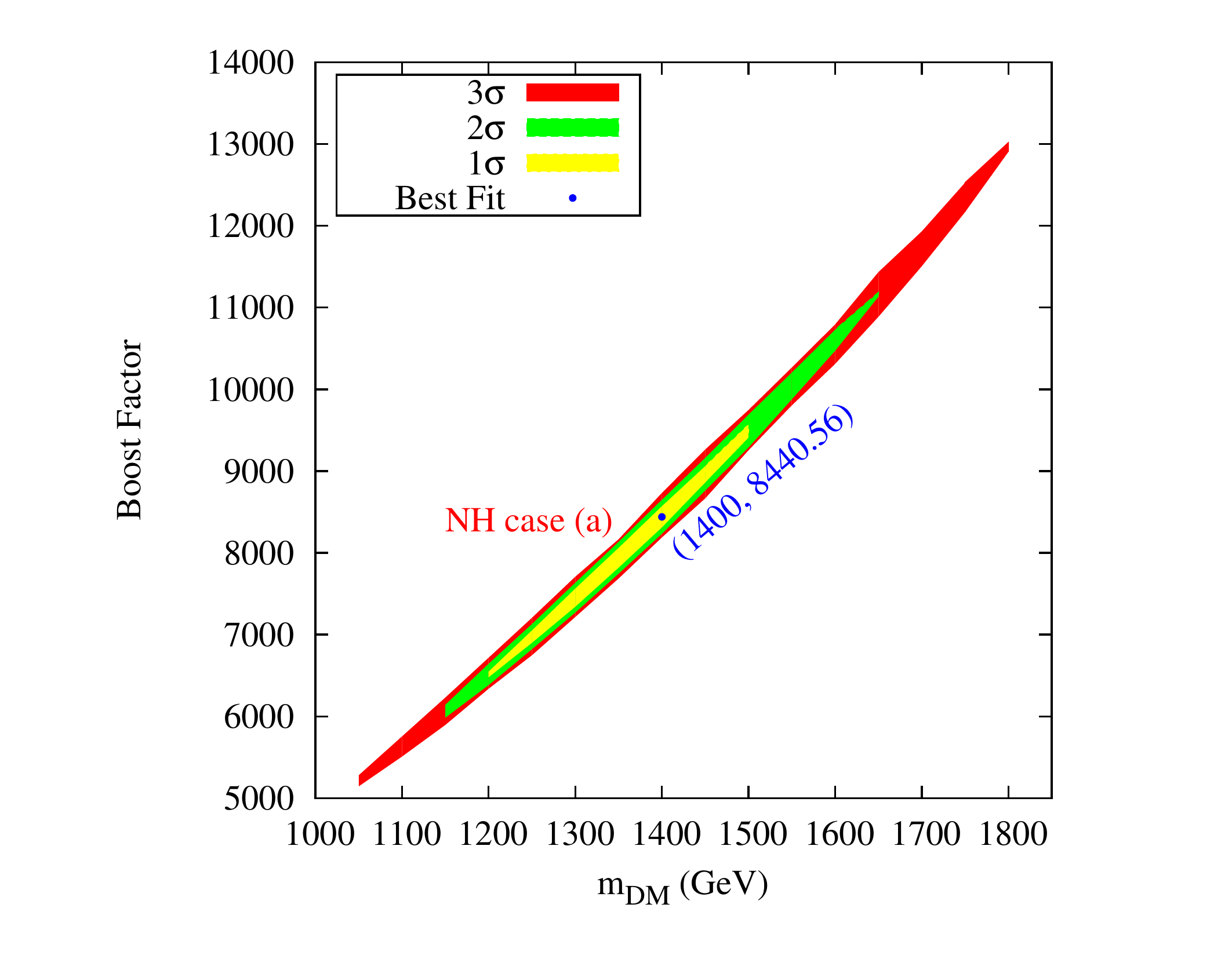}
\hspace{0.5cm}
\includegraphics[width=6.5cm]{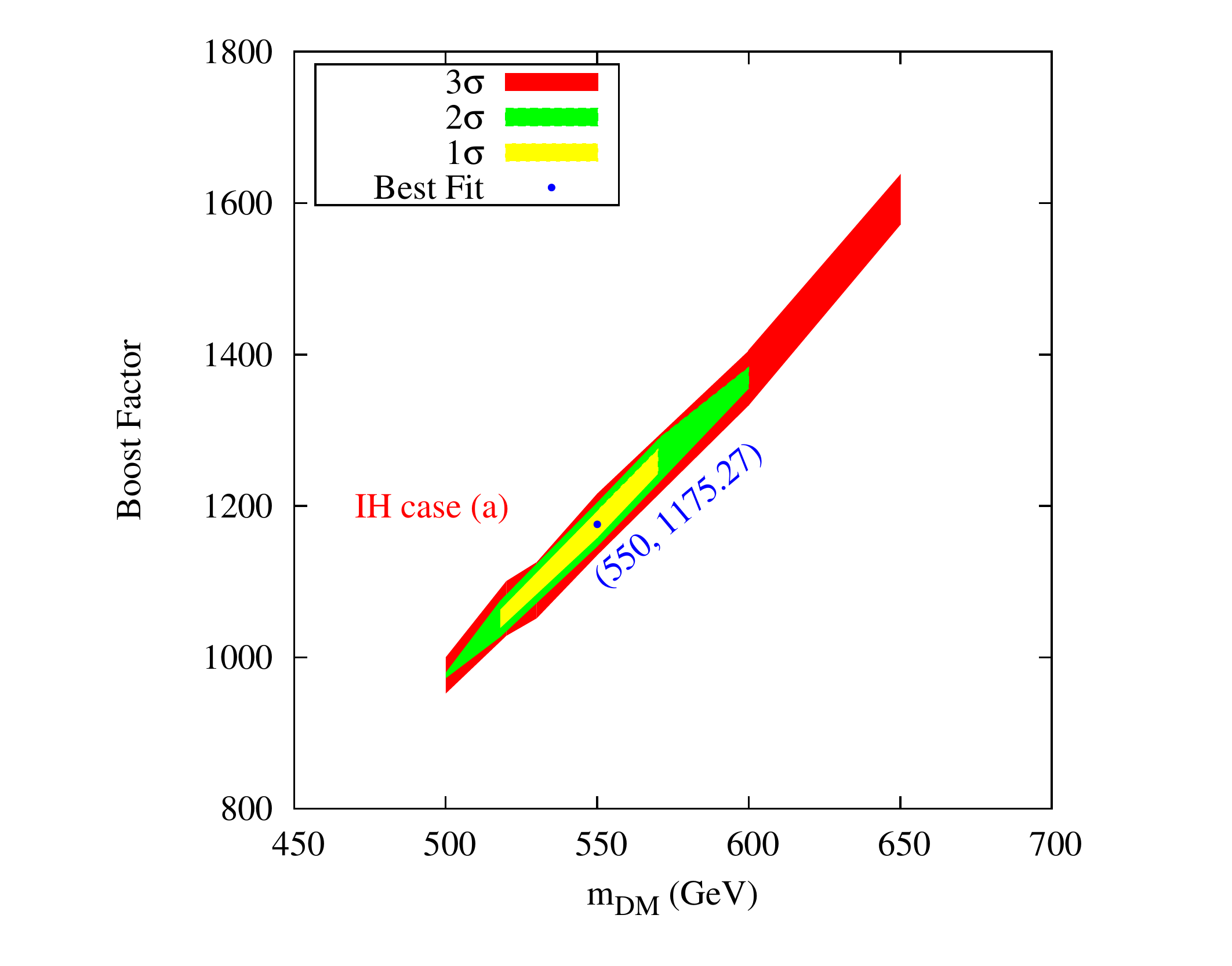}\\
\includegraphics[width=6.5cm]{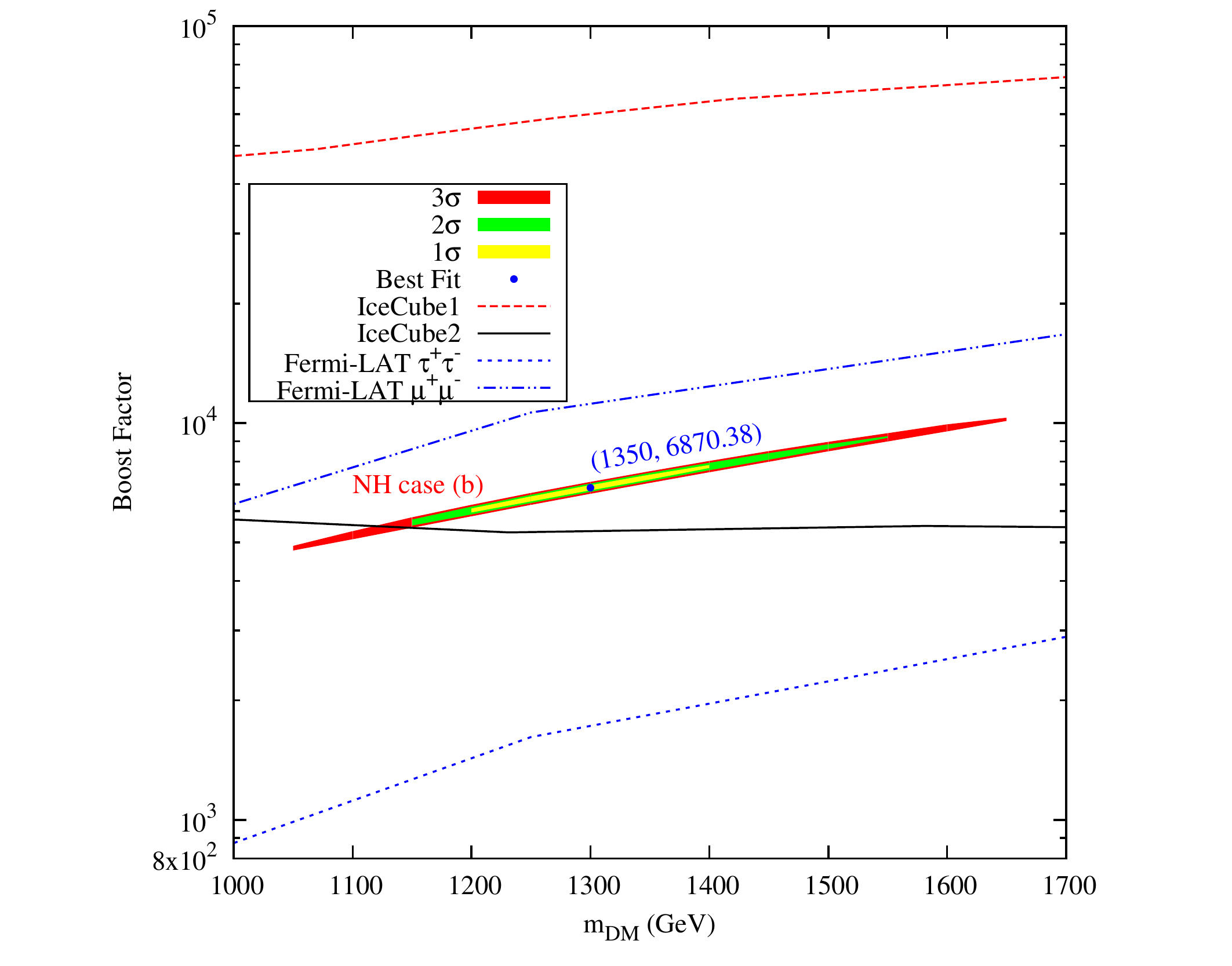}
\hspace{0.5cm}
\includegraphics[width=6.5cm]{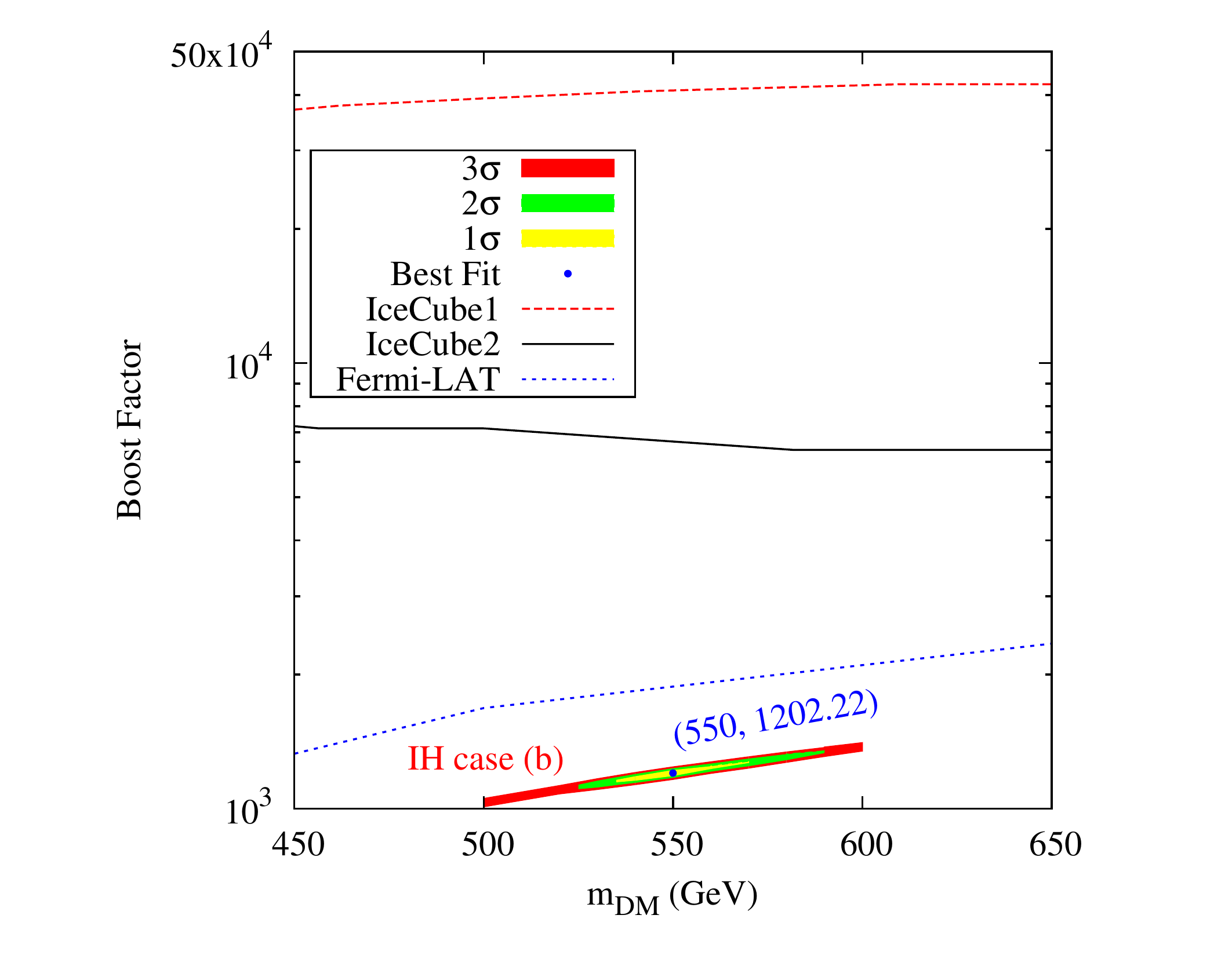}
\caption{The best-fit values and the $1\sigma$, $2\sigma$ and $3\sigma$ preferred ranges of the DM mass and the boost factor in our model to explain the AMS-02 positron excess. The results are shown for both NH and IH, and for both cases (a) and (b). Our estimated upper limits on the boost factor derived 
from the IceCube limits on the DM annihilation rate in the galactic center~\cite{ic2} (IceCube1) and galactic halo~\cite{ic1} (IceCube2) as well as from the Fermi-LAT combined limit on the DM annihilation into leptonic final states 
in nearby dwarf galaxies~\cite{Ackermann:2013yva} are also shown for   
comparison in case (b) (lower panels).}
\label{fig4}
\end{figure}

Using the $\chi^2$-fit results from Section~\ref{sec:3a}, we obtain the $1\sigma$, $2\sigma$ and $3\sigma$ allowed ranges of the DM mass and boost factor in our model to explain the AMS-02 positron excess. This is shown in Fig.~\ref{fig4} for both NH and IH cases, and also for both cases (a) and (b). For each case, we also show the best-fit point with minimum $\chi^2$ value. We find that the NH case prefers slightly 
higher values of the DM mass and boost factor, compared to the IH case. 

The large boost factors as shown in Fig.~\ref{fig4} 
to explain the AMS-02 data will also lead to an associated large gamma-ray 
flux due to inverse Compton scattering and bremsstrahlung effects caused by 
the relativistic leptonic final states. Moreover, our model Lagrangian [cf. Eq.~(\ref{lag})] being $SU(2)_L$-invariant, will also give rise to a large 
neutrino flux along with an excess positron flux in the DM annihilation. It is however important to observe that we have {\it four}-body final states due to the DM-pair annihilating to a pair of triplet scalars each of which then decay to a pair of leptons. Hence, it is in general difficult to translate the experimental upper bounds on the 
boost factor, derived from observations of gamma-ray flux and neutrino flux, since these bounds are usually obtained assuming the DM annihilation directly into {\it two}-body final states. We can make a rough comparison only in the limiting case (b) with DM and triplets are close to being mass-degenerate, such that the intermediate scalar triplets are produced almost at rest, and then decay to the leptonic final states. Due to the $SU(2)_L$-invariance, the branching ratio to charged lepton and neutrino final states will be 50\% each, as can also be verified from 
Eq.~(\ref{dm2}) for degenerate triplet scalars. Using this information, the 
annihilation cross section limits for neutrino final states, as given by IceCube from searches for DM annihilation in the galactic center~\cite{ic2} 
(IceCube1) and in the galactic halo~\cite{ic1} (IceCube2), have been translated to upper limits on boost factor in Fig.~\ref{fig4} (lower panels). 
Here we have taken the thermal relic abundance to be $\langle \sigma v\rangle_{\rm th}=2.2\times 10^{-26}~{\rm cm}^3{\rm s}^{-1}$~\cite{relic}, and have assumed the same boost factor for the annihilation into positrons and neutrinos. 
We do not show the corresponding IceCube limits due to other final states since these are weaker compared to that from the neutrino final states. Also we do 
not show here the IceCube limit from DM annihilation in nearby galaxy clusters~\cite{ic3} although it is slightly stronger than the corresponding limit from galactic halo~\cite{ic1}. The reason is that the limit for neutrino final states 
quoted in~\cite{ic3} was derived for Virgo cluster with sub-halos, and may not be a generic bound applicable to local (galactic) isotropic sources of DM.

Similar limits on the boost factor can be derived from the searches for 
gamma-rays due to DM annihilation. Here we use the Fermi-LAT combined 
limits on the DM annihilation rate from the diffuse gamma-ray search in nearby dwarf spheroidal galaxies~\cite{Ackermann:2013yva}, and translate them into the upper limits on the boost factor for our case (b), as shown in Fig.~\ref{fig4} 
(lower panels). For the NH case, we expect mostly muon and tau final states, each occurring with roughly 50\% branching ratio, whereas for the IH case, we expect mostly electron final states. Hence, we have used the corresponding 
Fermi-LAT limits for muon and tau final states in the NH case, and for electron final state in the IH case. 

From Fig.~\ref{fig4} we conclude that the boost-factors required to explain the 
AMS-02 data in our model are still consistent with the experimental limits from observations of neutrino and gamma-ray fluxes, except for the NH case (b), which is ruled out by the Fermi-LAT limit on the $\tau^+\tau^-$ channel.  However, 
we must keep in mind that there are some inherent astrophysical 
uncertainties in our derived 
boost factor limits, e.g. halo uncertainties in the DM density profile. Also the boost factor depends on the distance from the halo center, and a positron boost factor may not be the same as that for neutrinos/gamma-rays, 
as their propagation is not the same.  
For instance, a nearby DM clump would tend to increase the positron flux as positrons have much shorter emission range (of order of 100 kpc) due to their rapid energy loss~\cite{Moskalenko:1997gh}, whereas the neutrino/gamma-ray flux 
could also get contributions from distant sources. Hence, our comparison of the positron boost factors with the experimental limits derived from neutrino and gamma-ray fluxes is just a rough estimate, and we could easily gain up to an order of magnitude due to these inherent astrophysical uncertainties. Moreover, these limits are in general not applicable to our case (a) when the triplets are 
highly boosted, and the leptonic energy spectrum due to our 4-body final state 
will be different from that of the 2-body final state for a given DM-mass.
 Nevertheless, improved limits from future IceCube and Fermi-LAT 
data should be able to completely rule out this model in the absence of a DM 
signal.

It is important to mention here that the large boost factors considered in Fig.~\ref{fig4} could also lead to observable effects in the BBN and CMB precision measurements. Although our boost factors are well within the 
upper limit derived from the BBN data on light element abundances~\cite{Hisano:2009rc}, the corresponding limits derived from the CMB power spectrum~\cite{Cline:2013fm, Madhavacheril:2013cna} are potentially in conflict with our best-fit values. 
In particular, the CMB bounds are only sensitive to the total electromagnetic 
energy injected to the primordial plasma due to DM annihilation 
after recombination (corresponding to a redshift $z\approx 1100$), and hence, 
equally applicable to both 2-body and 4-body leptonic final states.  
However, one should note that these CMB limits on the DM annihilation rate become weaker at lower redshift values, as the fraction of energy absorbed by the CMB plasma is small for $z\lsim 100$~\cite{Madhavacheril:2013cna}. Thus, if the large boost factors, as required today ($z=0$) to explain the AMS-02 data, are partly/entirely due to a clumpy DM substructure in the galactic halo, which grows at lower redshifts ($z<100$) for typical halo mass functions~\cite{Huetsi:2009ex}, one can successfully avoid the CMB limits. On the other hand, for an entirely particle physics origin, in which case the large boost factors must be present even at larger redshifts, it is difficult to avoid the stringent limits from CMB. It is however possible that the Breit-Wigner mechanism mentioned earlier in Section~\ref{sec3} could still account for a fraction of the enhancement in the boost factor, as much as allowed by the CMB constraint, with the remaining enhancement coming from an 
astrophysical origin. In addition, it might be possible to fit the AMS-02 positron rise with a smaller (or even no) boost factor, if the DM candidate had a non-thermal cosmological history, so that the annihilation rate could be larger than the thermal annihilation rate, while still satisfying the relic density constraint (see e.g.,~\cite{Gelmini:2006pw, Easther:2013nga}).  
\section{Predictions for Fluxes}\label{sec4}
\begin{figure}[h!]
\centering
\includegraphics[width=6.4cm]{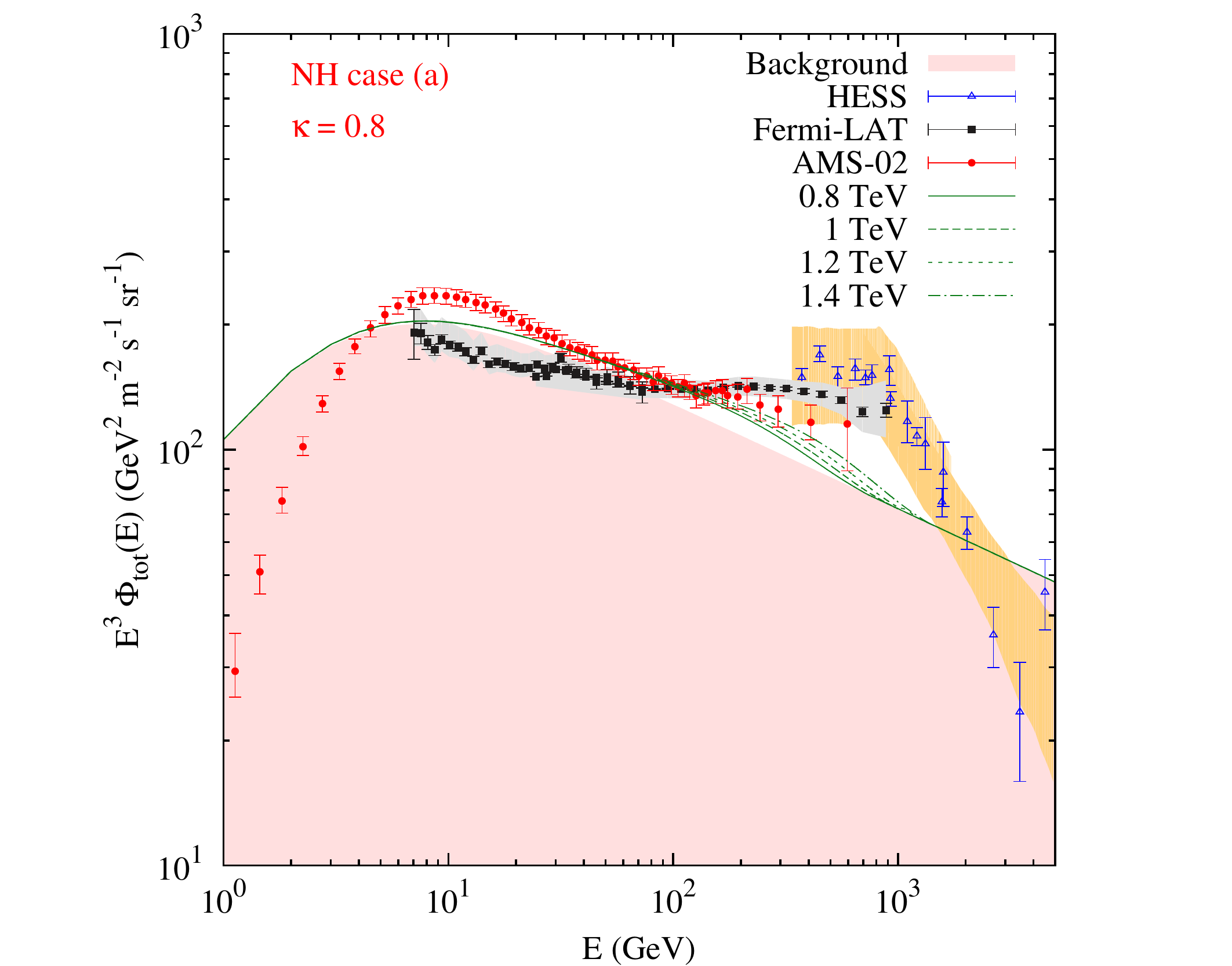}
\hspace{0.5cm}
\includegraphics[width=6.4cm]{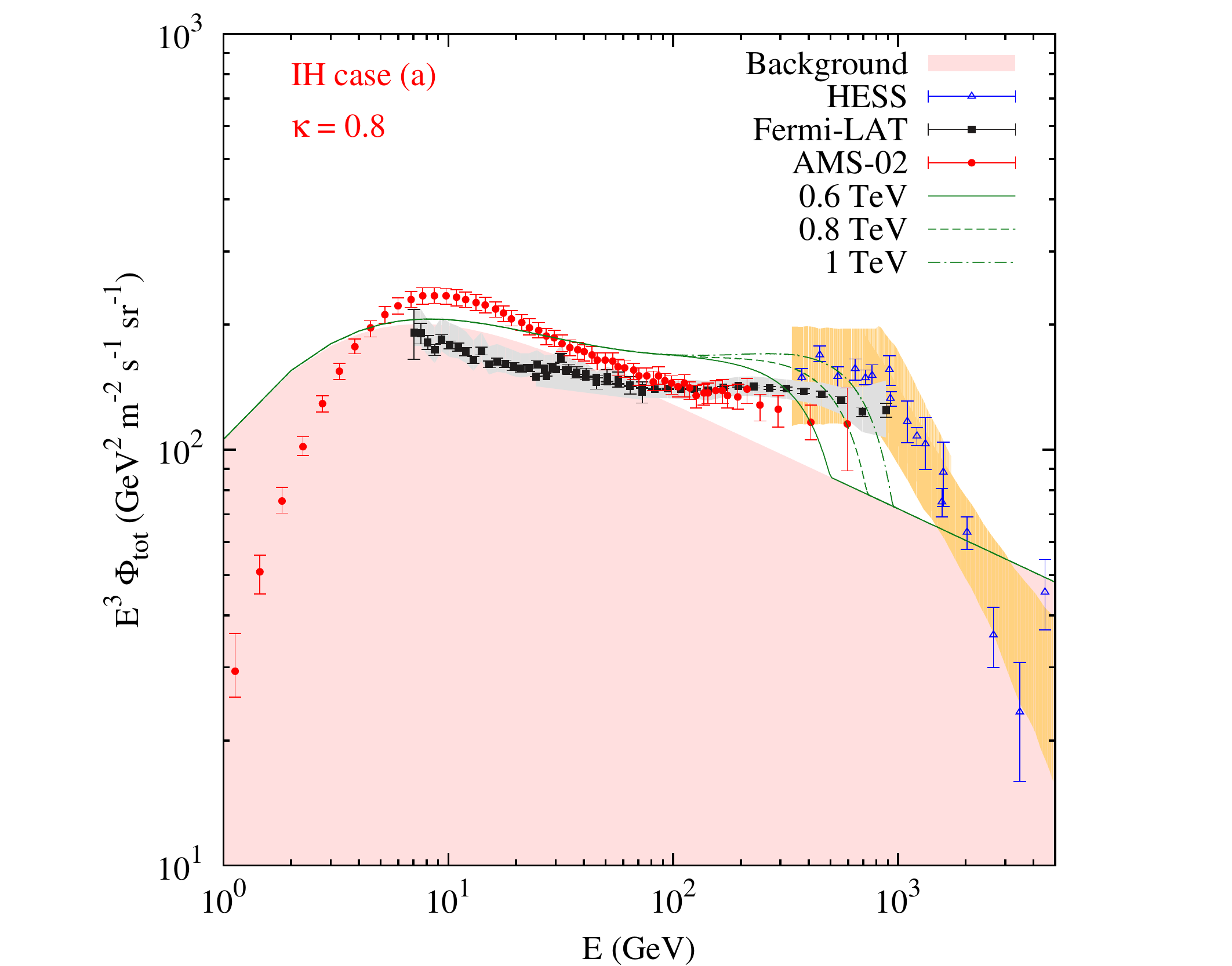}\\
\includegraphics[width=6.4cm]{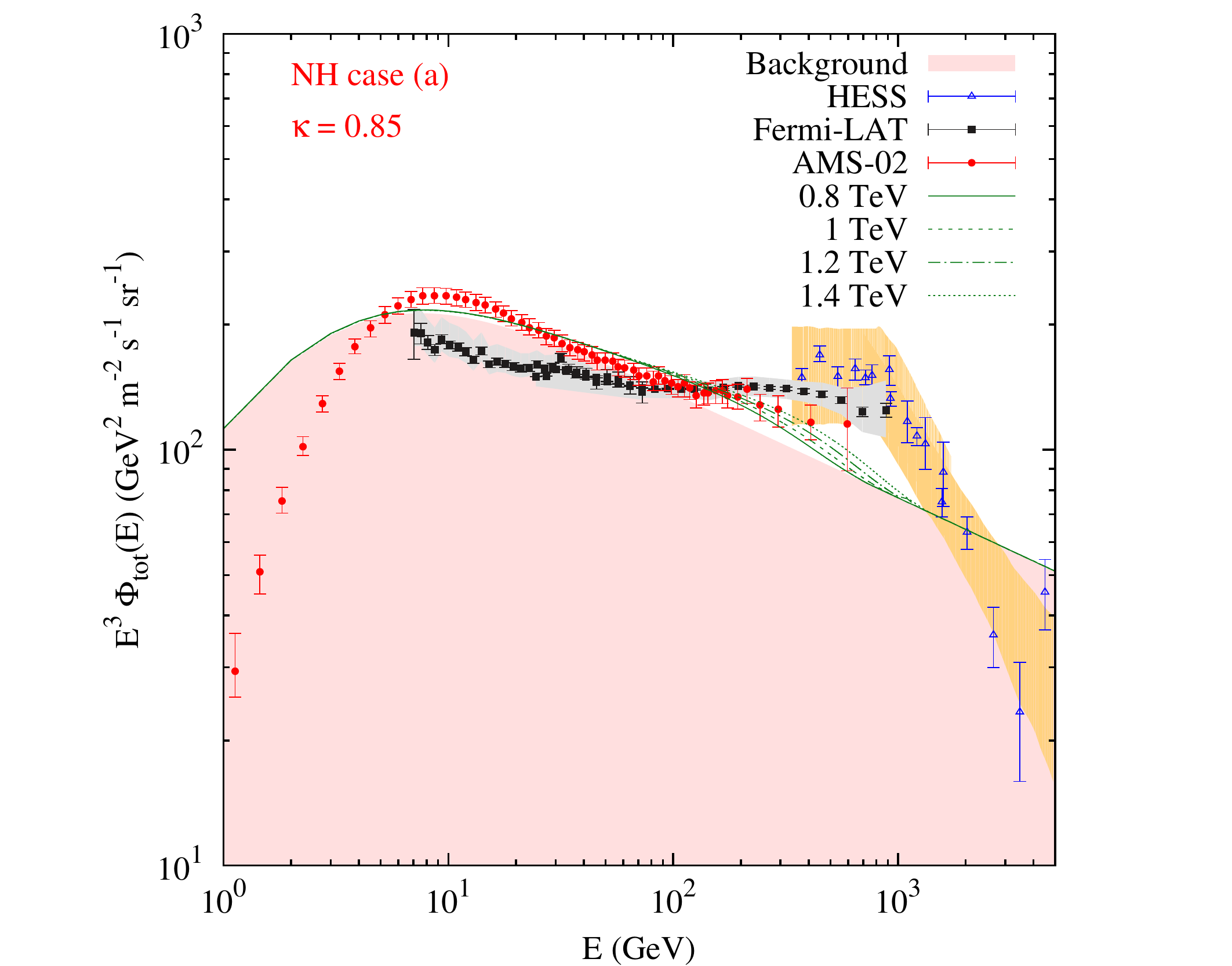}
\hspace{0.5cm}
\includegraphics[width=6.4cm]{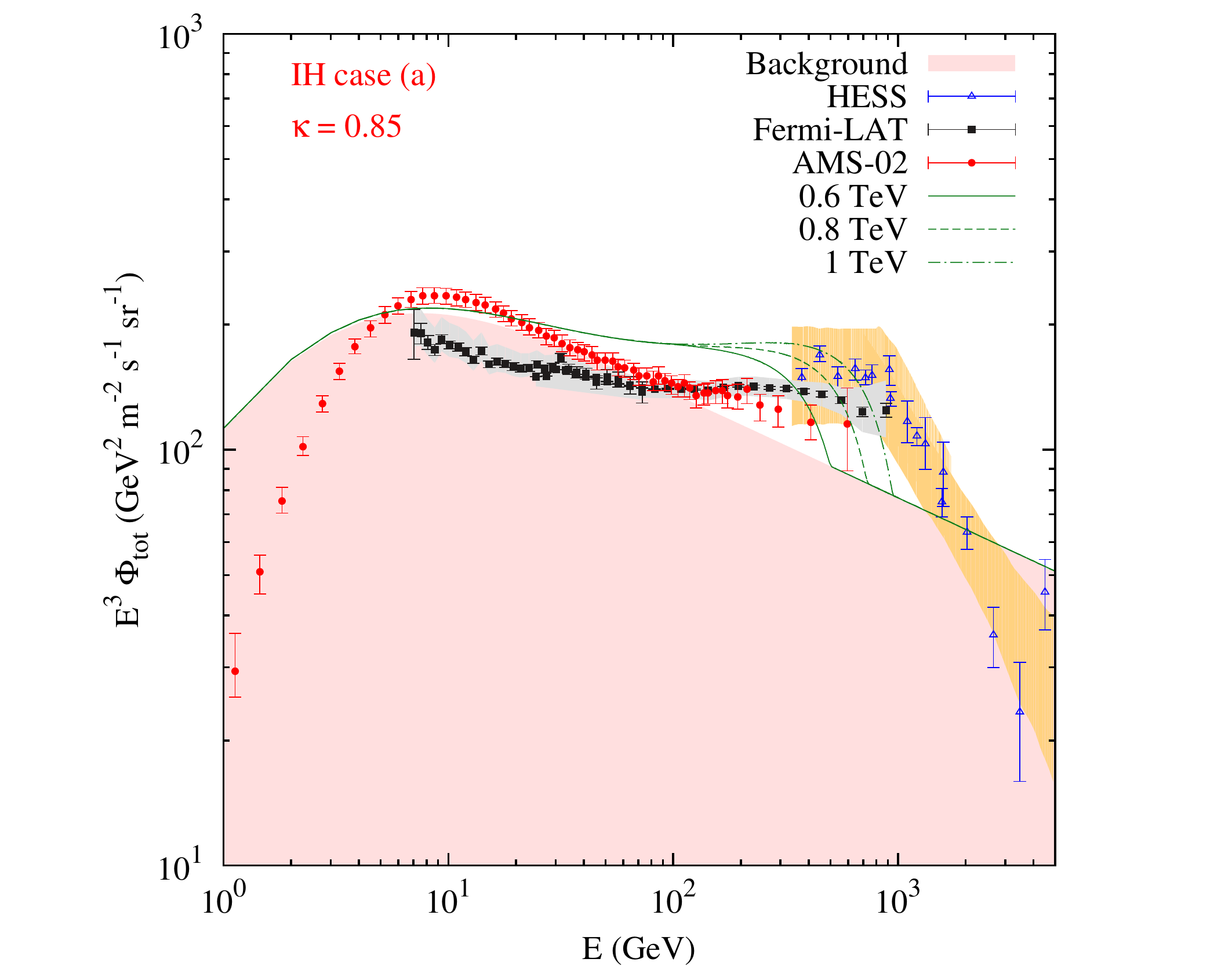}\\
\includegraphics[width=6.4cm]{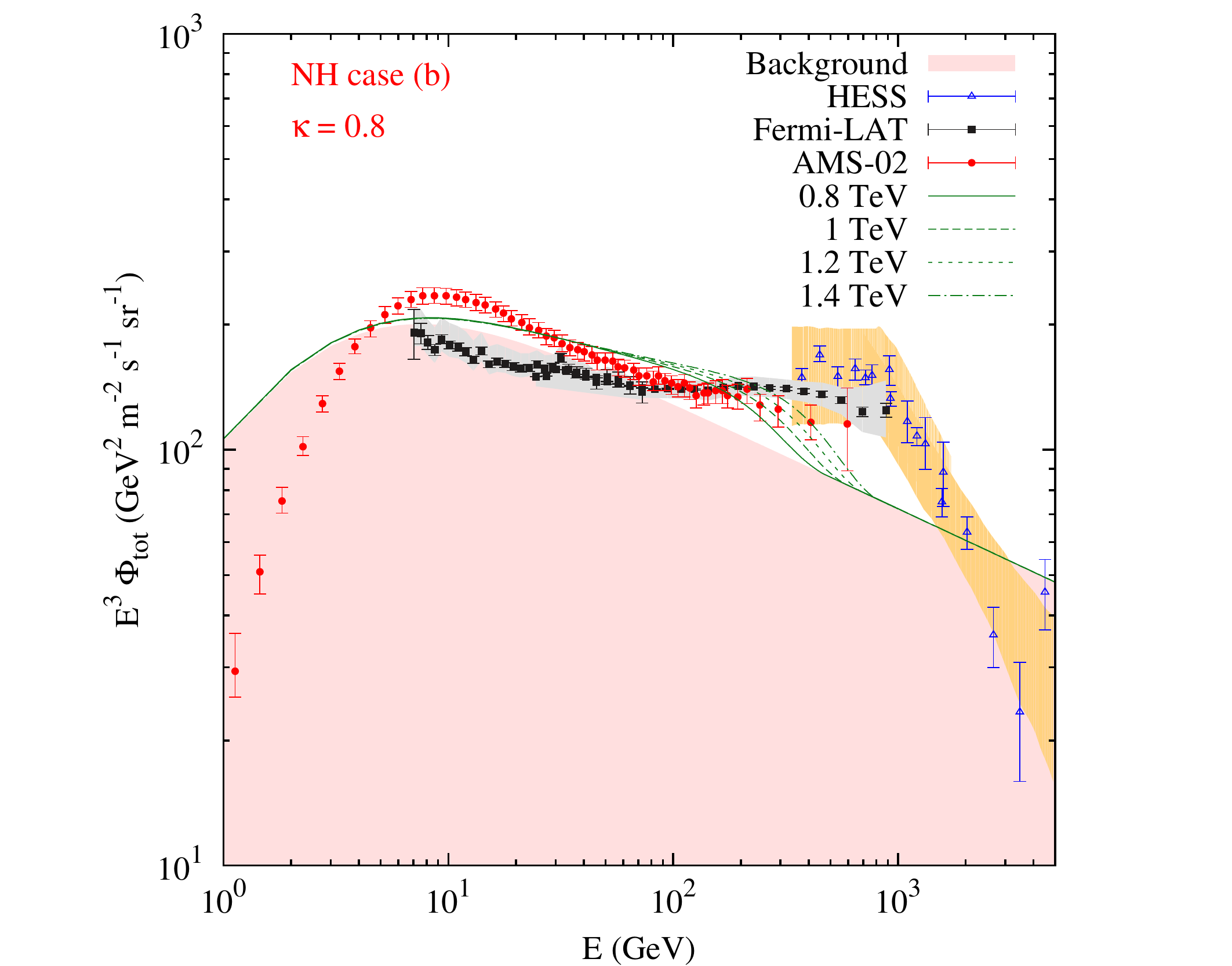}
\hspace{0.5cm}
\includegraphics[width=6.4cm]{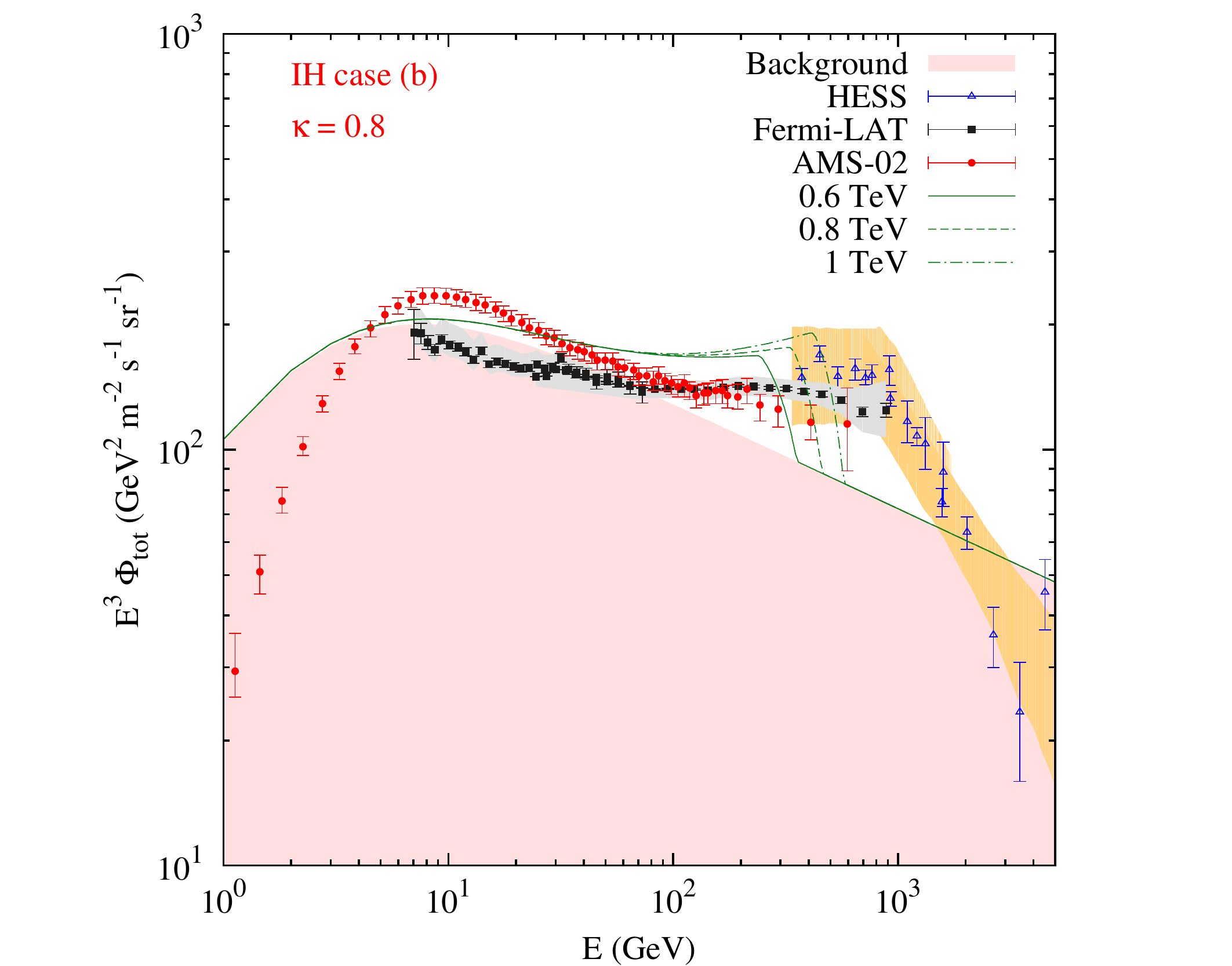}\\
\includegraphics[width=6.4cm]{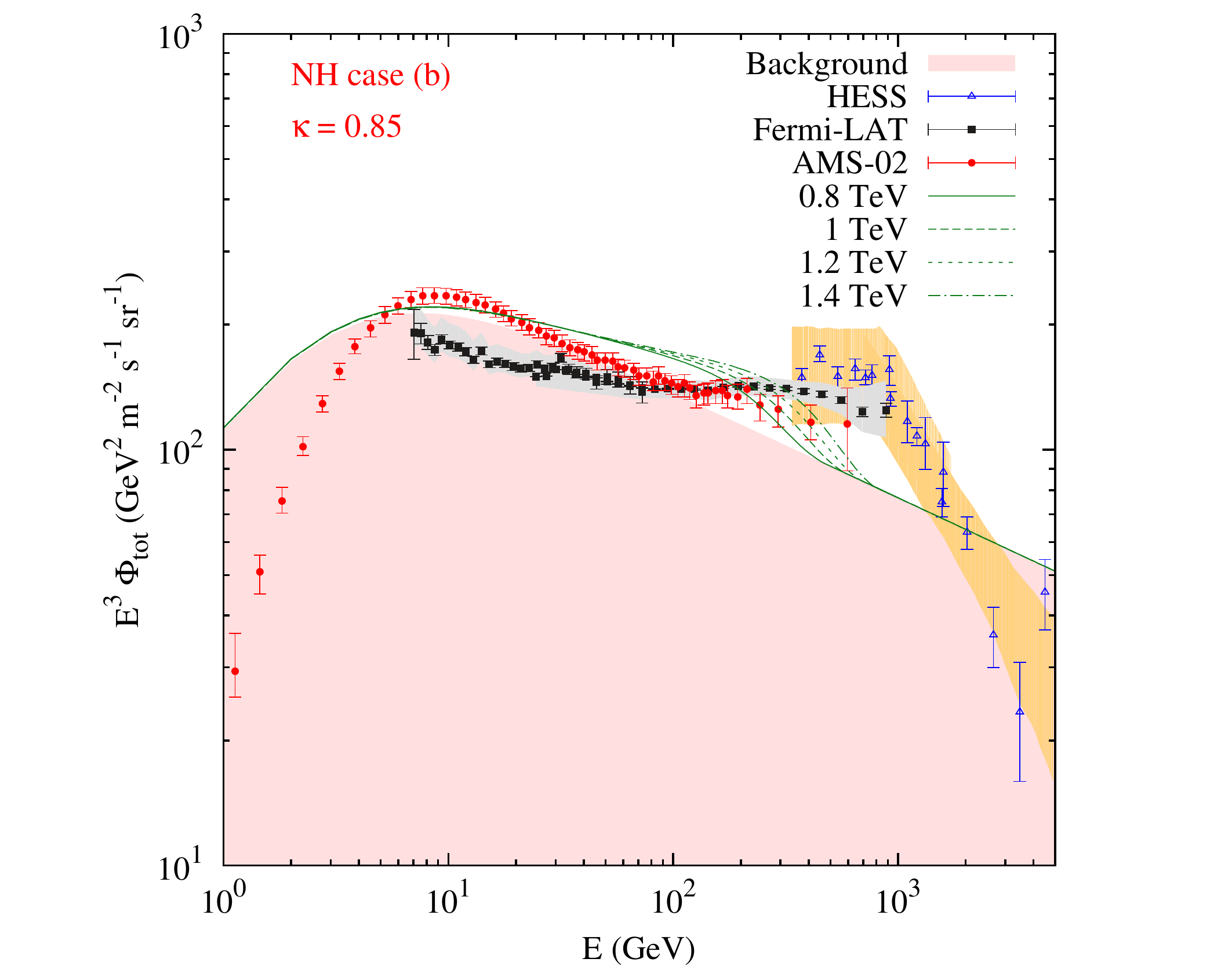}
\hspace{0.5cm}
\includegraphics[width=6.4cm]{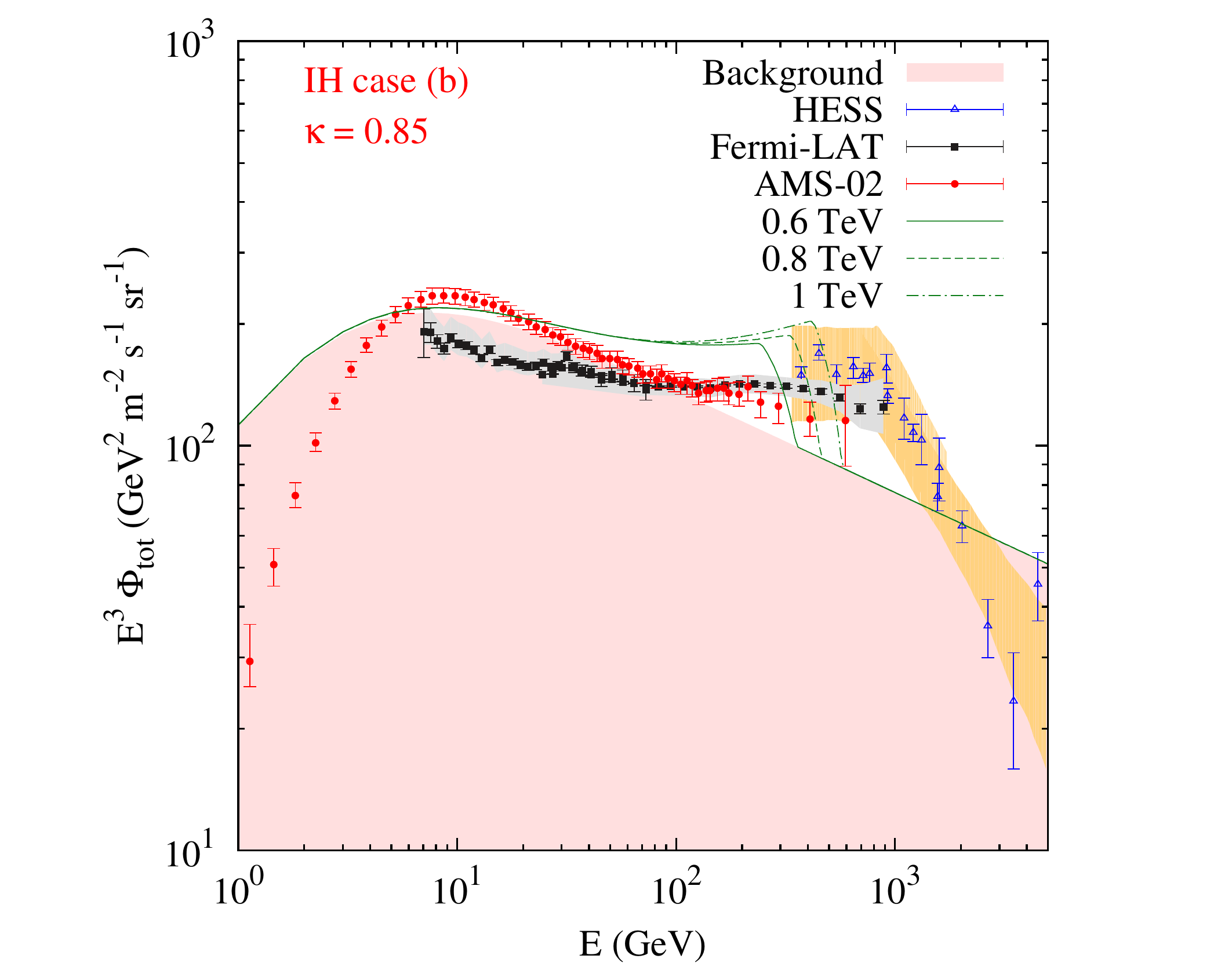}
\caption{The $e^-+e^+$ spectrum ((energy)$^3\times$ total flux) in our model for both NH (left panel) and IH (right panel) in two kinematically distinct cases (a) and (b) with various DM masses (in TeV), and with $\kappa=0.8,~ 0.85$. 
Also shown are the AMS-02~\cite{AMS-update}, Fermi-LAT~\cite{fermi-flux2} and HESS~\cite{hess2} data for the total flux (with error bars), and the total background flux from cosmic rays (shaded region) for comparison. 
}
\label{fig2}
\end{figure}
Apart from explaining the AMS-02 positron fraction data, the model predictions for the positron flux and the total electron+positron flux are also compatible with the recent AMS-02 flux measurements~\cite{AMS-update}. Our results for the total electron+positron flux obtained from Eq.~(\ref{phitot}) are shown in Fig.~\ref{fig2} for two benchmark values of $\kappa=0.8$ and 0.85. We show our results for both cases (a) and (b) and for both NH and IH scenarios. The slight excess above the background flux at higher energies can be explained by our DM signal for $\kappa=0.80$ - 0.85 for NH case (a), while the fluxes for NH case (b), and for IH both cases (a) and (b), are much larger than the observed values, and hence, disfavored. 

For comparison, we also show the total cosmic electron+positron flux obtained by Fermi-LAT~\cite{fermi-flux2} in the energy range of 10 GeV--1 TeV and by HESS~\cite{hess2} in the high energy regime. The systematic errors in the Fermi-LAT data are shown by the gray band, while those in the HESS data are shown by the orange band. Note that for $E>100$ GeV, the AMS-02 flux values are {\it lower} compared to the Fermi-LAT and HESS values. Thus, in a generic DM annihilation model, where the excess electron+positron flux is expected to show a drop in the spectrum, along with the excess positron fraction, at higher energies, as we approach the DM mass, a fit to the AMS-02 data alone will prefer a {\it lower} value of the DM mass, whereas a global fit including Fermi-LAT and/or HESS data sets, which show a fall-off behavior at much higher energies, will prefer a larger DM mass. This apparent tension between the AMS-02 spectrum and other data sets makes a simultaneous fit rather difficult~\cite{DM1}.

\begin{figure}[t!]
\centering
\includegraphics[width=6.4cm]{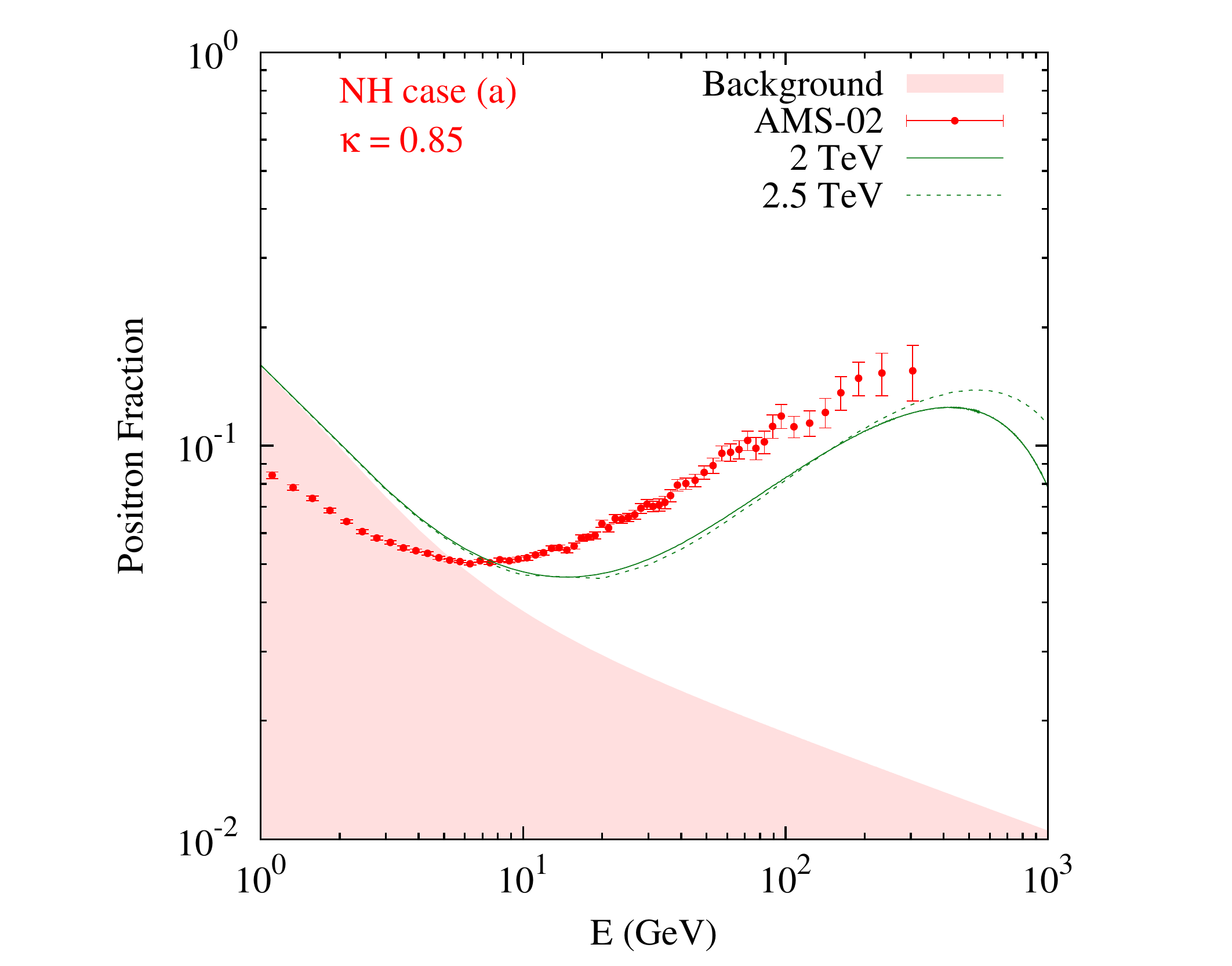}
\hspace{0.5cm}
\includegraphics[width=6.4cm]{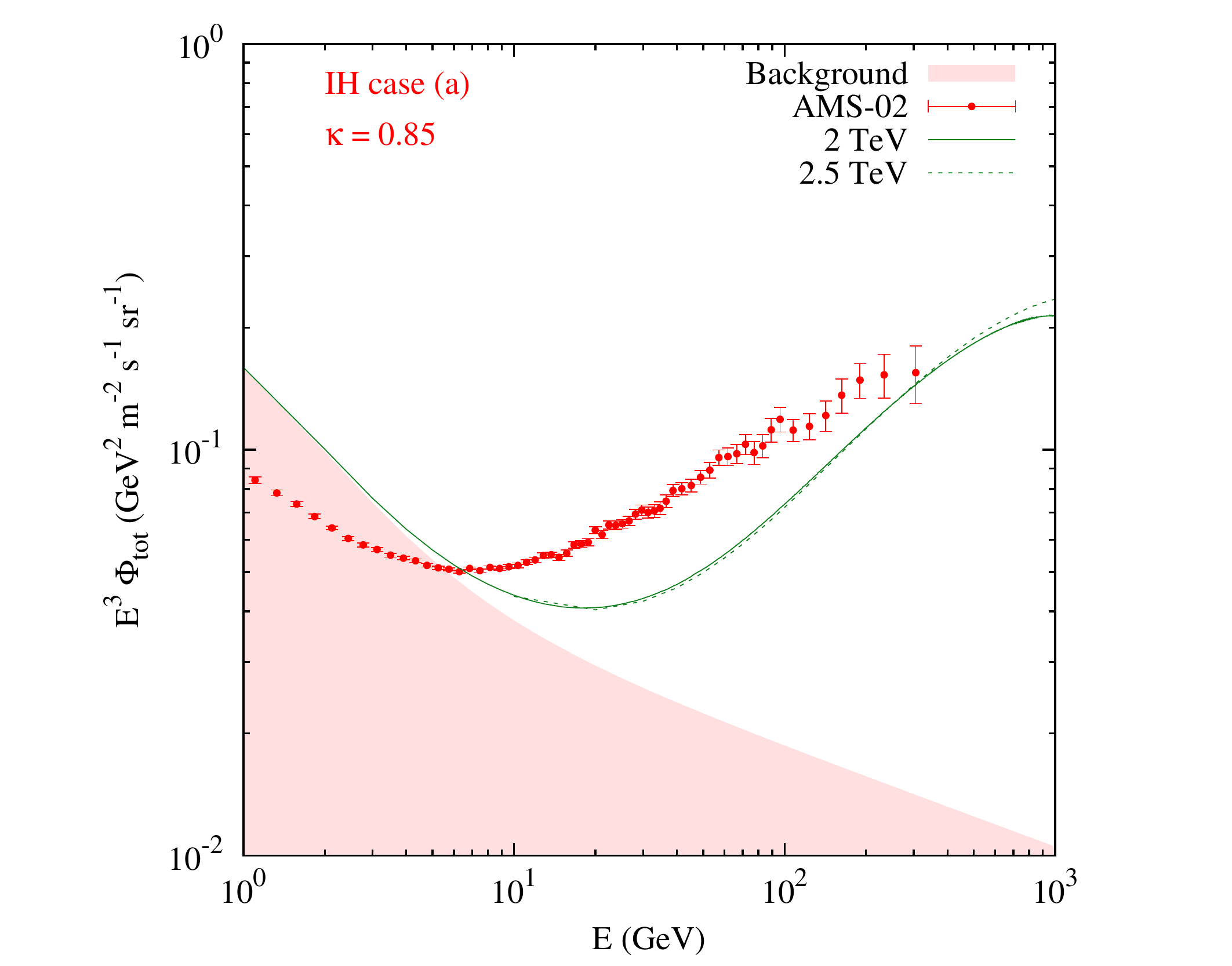}\\
\includegraphics[width=6.4cm]{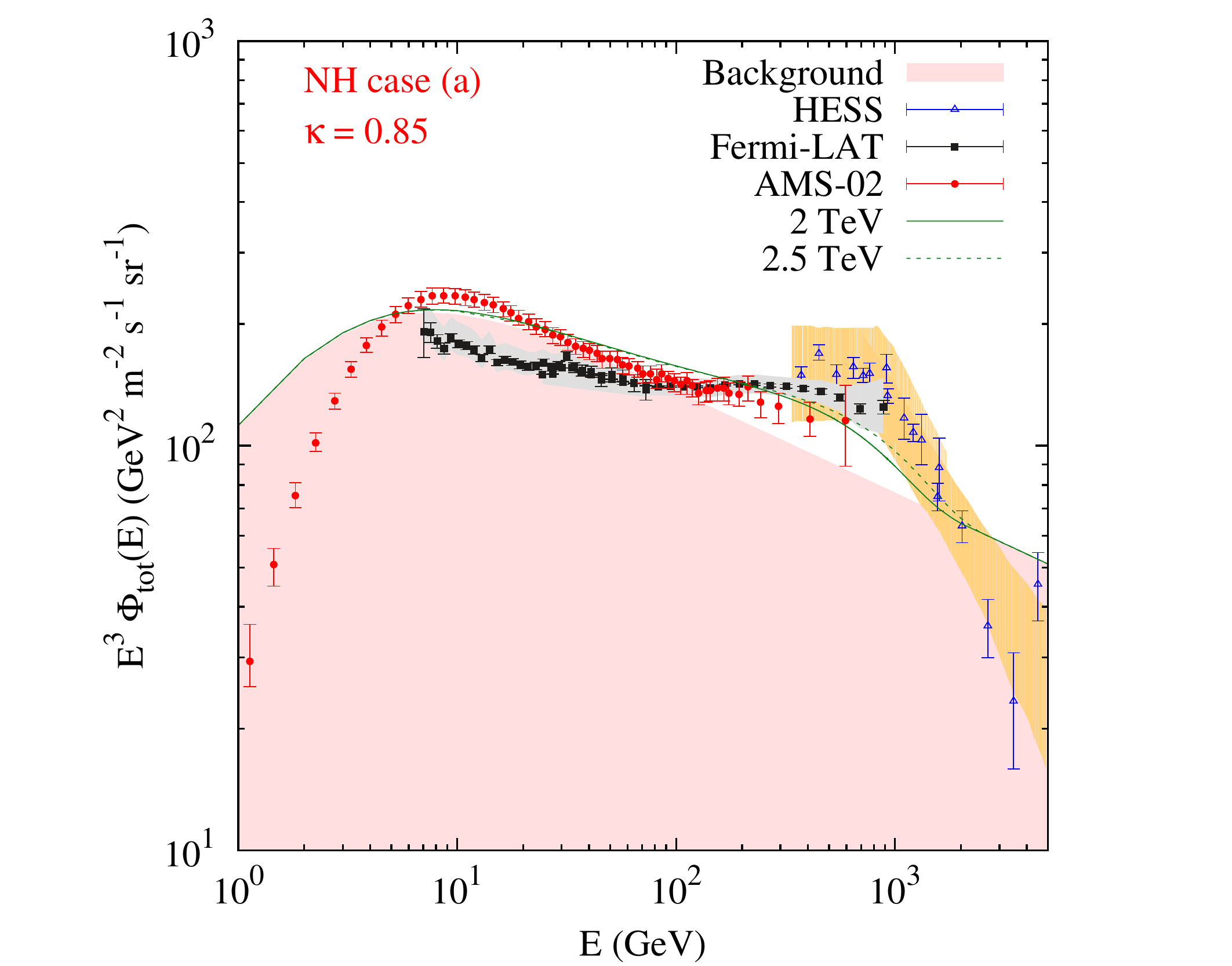}
\hspace{0.5cm}
\includegraphics[width=6.4cm]{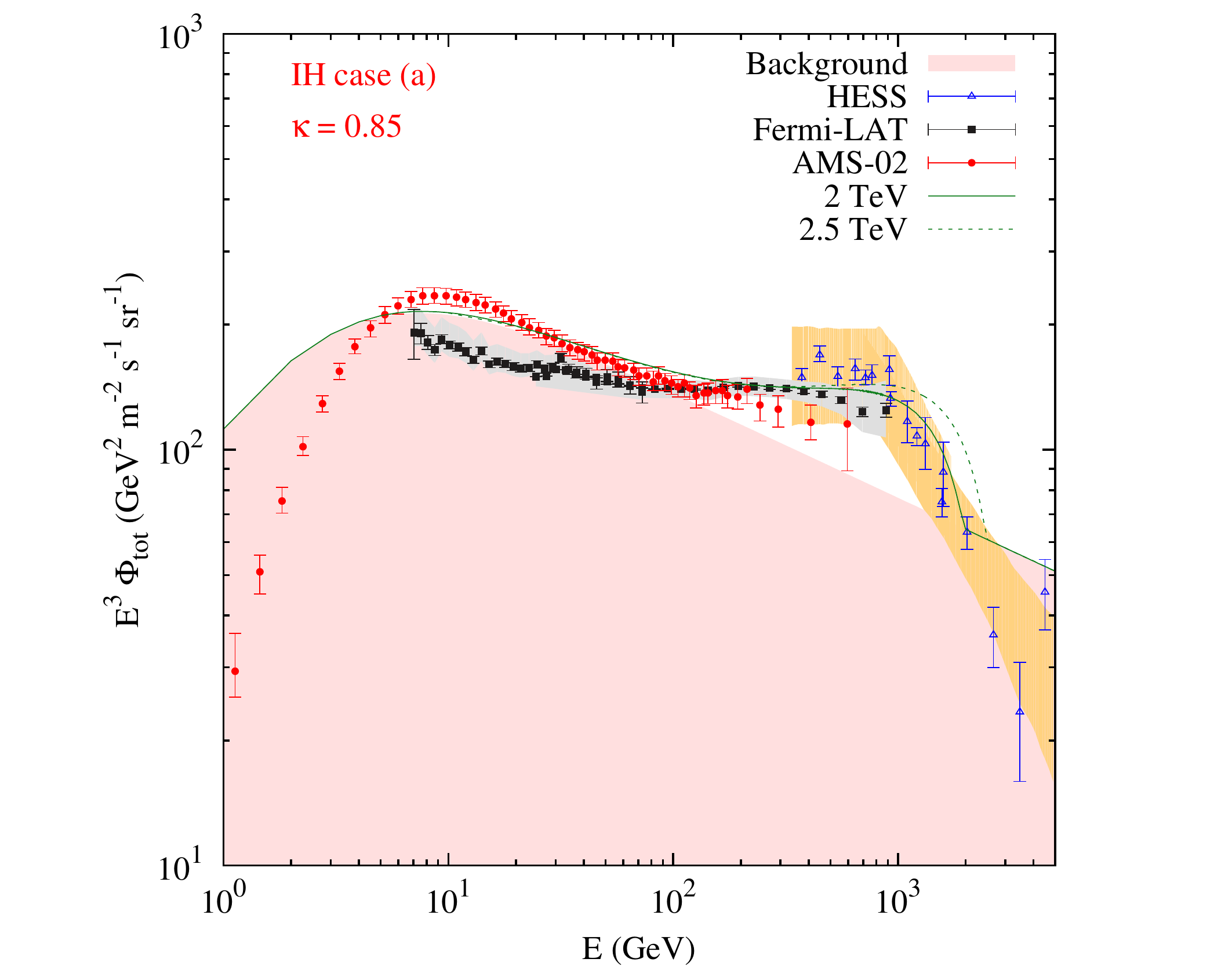}
\caption{A combined $\chi^2$-fit to the AMS-02 positron fraction and Fermi-LAT $e^-+e^+$ spectrum in our model for NH (left panel) and IH (right panel). Here we have chosen case (a) $m_\Delta^2\ll m_{\rm DM}^2$ and $\kappa=0.85$, and have considered two benchmark values of DM masses for illustration. }
\label{fig3a}
\end{figure}

To illustrate this discrepancy between different experimental results, we perform a combined $\chi^2$-fit to the AMS-02 positron fraction and the Fermi-LAT $e^-+e^+$ spectrum. The results are shown in Figure~\ref{fig3a} for both NH and IH. Here we have chosen $m_\Delta^2\ll m_{\rm DM}^2$ [case (a)] and $\kappa=0.85$ in Eq.~(\ref{phitot}) for the same background model as chosen for Figures~\ref{fig1} and \ref{fig2}. We have considered two benchmark values of $m_{\rm DM} = 2$ and 2.5 TeV for illustration. Lower values of DM mass will lead to a better fit to the AMS-02 positron fraction data (cf. Figure~\ref{fig1}), but worsen the fit to the Fermi $e^-+e^+$ spectrum (cf. Figure~\ref{fig2}). On the other hand, higher values of DM mass give a better fit to the Fermi spectrum, whereas the AMS-02 positron fraction fit becomes unacceptable, as can already be seen from Figure~\ref{fig3a}. For completeness, we present in Table~\ref{tab3} the numerical values of the boost factor and $\chi^2$/dof for the benchmark points shown in Figure~\ref{fig3a}. The corresponding fits for case (b) will be worse, and hence, are not shown here. Also, including the HESS $e^-+e^+$ spectrum in the analysis will lead to a larger $\chi^2$-value. 
From Figure~\ref{fig3a} and Table~\ref{tab3}, it is clear that although it is possible to accommodate the Fermi-LAT (and also HESS) $e^-+e^+$ spectrum with a heavier DM, it becomes increasingly difficult to explain the AMS-02 positron fraction data.  Of course, it might be possible to improve the fit significantly with a different background model than that considered here, especially due to the fact that the Fermi-LAT measured flux in the low-energy regime (for $E<100$ GeV) is slightly below the predicted background flux. However, in our opinion, the flux discrepancy can be best resolved only after AMS-02 releases the high-precision flux measurements, which will be crucial to test the viability of a generic DM interpretation of the positron excess. With this note, we will not consider the combined $\chi^2$-fit in our subsequent discussion.    
\begin{table}[h!]
\begin{center}
\begin{tabular}{c|c|c|c}\hline\hline
Case & $m_{\rm DM}$ (TeV) & boost factor & $\chi^2_{\rm min}$/dof \\ \hline
NH & 2 & 8822.35  & 84.79\\
& 2.5 &  12167.8  & 88.3 \\ \hline
IH & 2 & 3225.75 & 112.71 \\
& 2.5 & 7275.12 & 116.45 \\ \hline
\hline
\end{tabular}
\end{center}
\caption{The best-fit values of the boost factor along with the corresponding $\chi^2$ per degrees of freedom (dof=48) for a combined $\chi^2$-analysis of AMS-02 positron fraction and Fermi-LAT $e^-+e^+$ spectrum. Here we have chosen $\kappa=0.85$ in Eq.~(\ref{frac}) and have only shown the results for case (a) $m^2_\Delta\ll m^2_{\rm DM}$.}
\label{tab3}
\end{table}

\begin{figure}
\centering
\includegraphics[width=6.5cm]{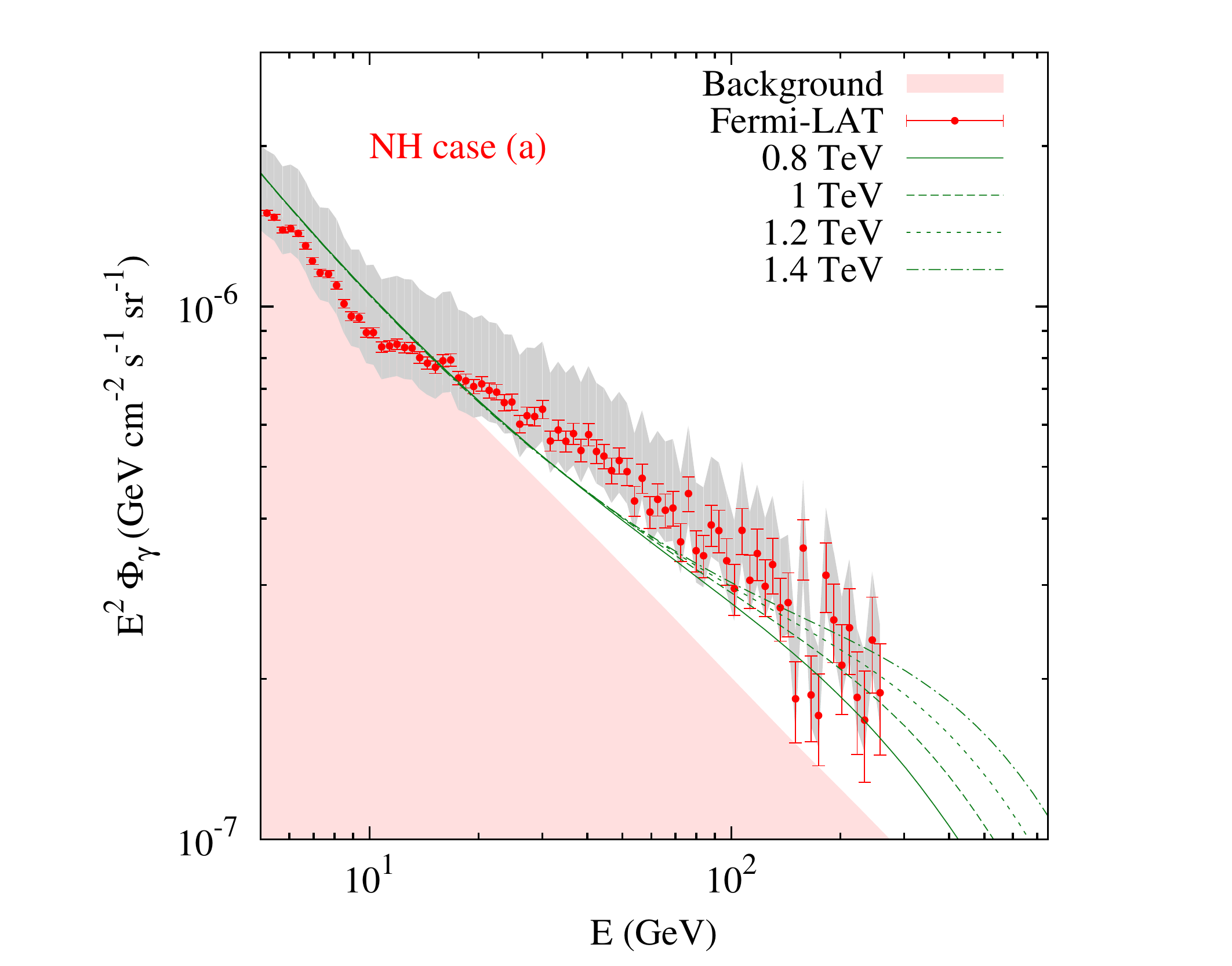}
\hspace{0.5cm}
\includegraphics[width=6.5cm]{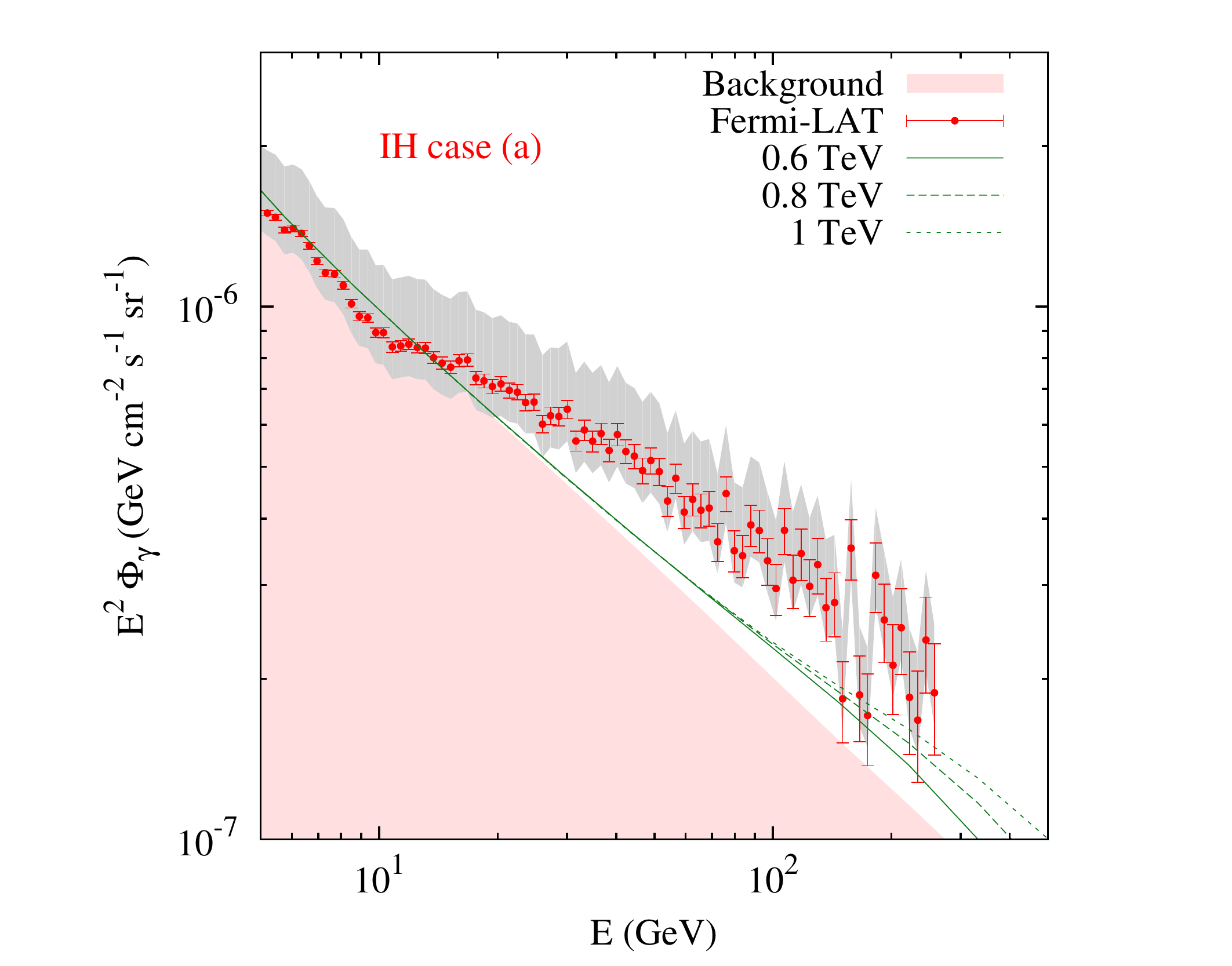}\\
\includegraphics[width=6.5cm]{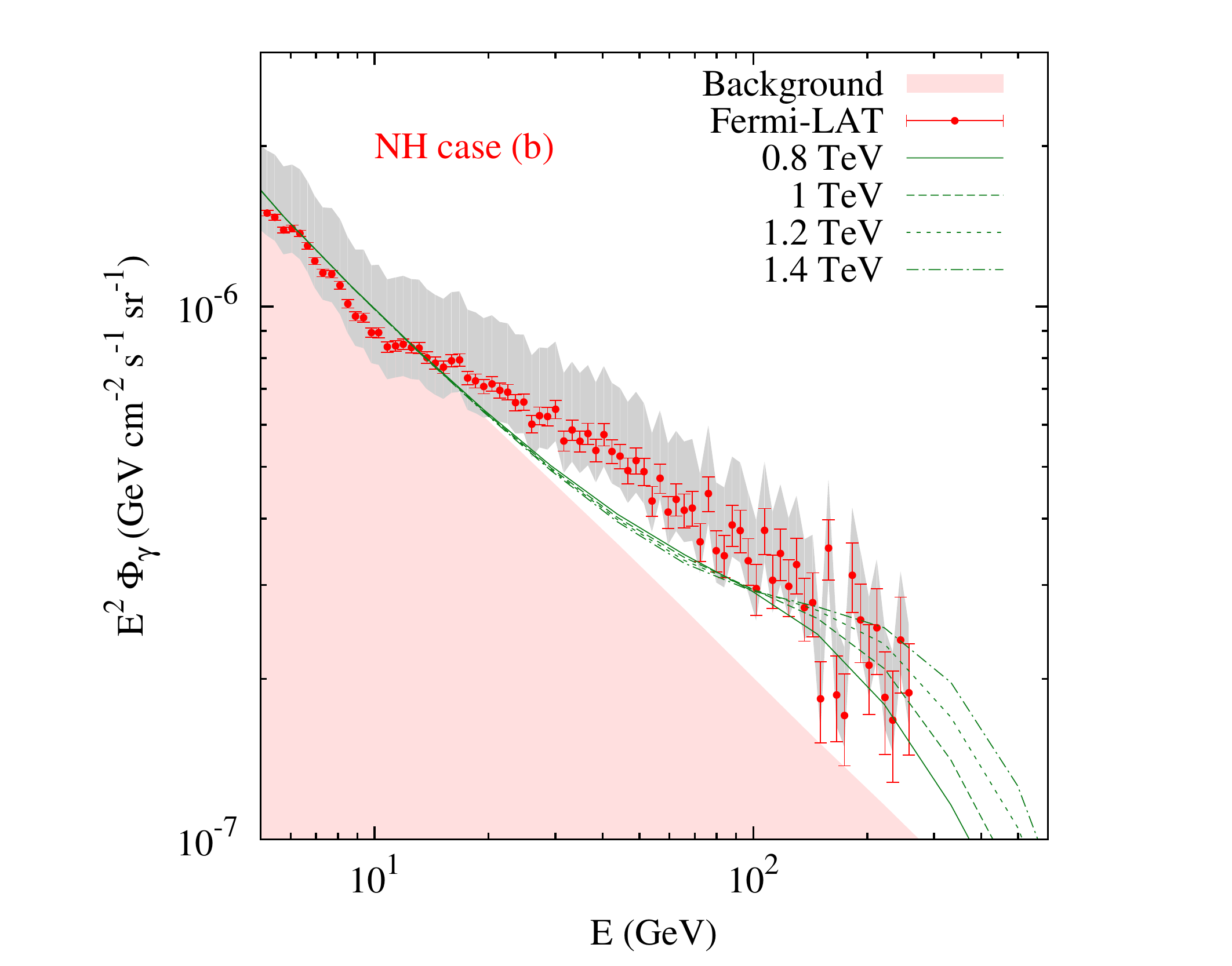}
\hspace{0.5cm}
\includegraphics[width=6.5cm]{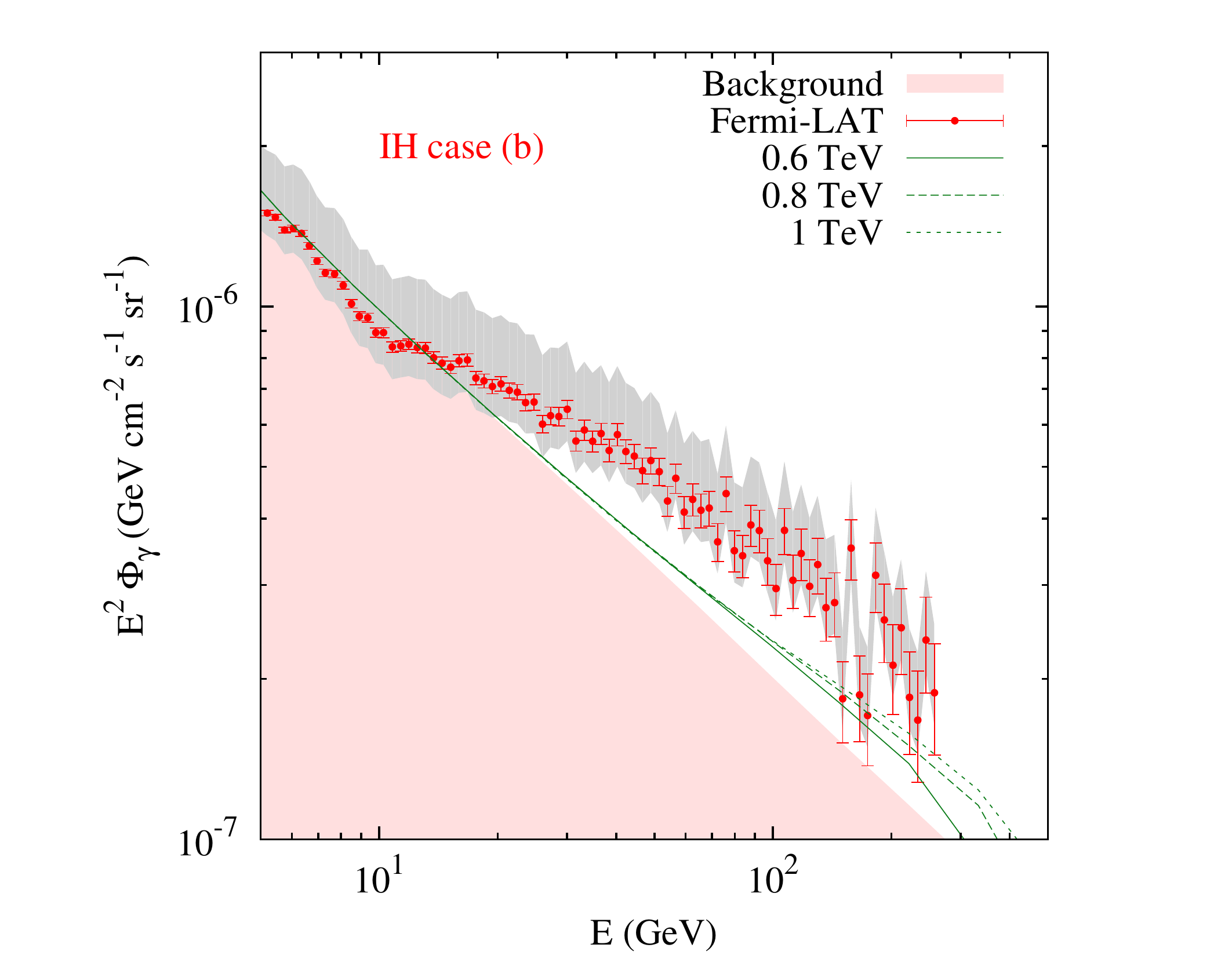}
\caption{The predictions for photon spectrum ((energy)$^2\times$ flux) in our model for both NH and IH in two distinct kinematic cases (a) and (b) with various DM masses (in TeV). The diffuse gamma ray background produced by cosmic rays (shaded region) and the Fermi-LAT measurements of the inclusive photon spectrum~\cite{fermi-gamma} are also shown. The gray band in the Fermi-LAT data shows the systematic error due to uncertainty in the effective area.}
\label{fig3}
\end{figure}

We also compute the photon flux due to DM annihilation in our model using {\tt micrOMEGAs},  and compare it with the diffuse background gamma-ray radiation produced by cosmic rays using {\tt GALPROP} and the Fermi-LAT observed inclusive continuum photon spectrum~\cite{fermi-gamma}. The results are shown in Fig.~\ref{fig3}.  For the  background gamma-ray radiation, we have included  three sources: inverse Compton scattering, bremsstrahlung and pion decay, averaged over the full sky. The gray band in the Fermi-LAT data shows the systematic error due to uncertainty in the effective area. The Fermi-LAT spectrum is for all sky minus (galactic plane+galactic center) and for high latitude 
region ($|b|>10^\circ$).\footnote{The bulk of galactic plane was not considered since the diffuse emission from interactions of cosmic rays with interstellar gas is strong in this region. The galactic center region was not considered since the DM signal from this region is strongly dependent on the uncertainty in the DM density, and only cuspy profiles lead to strong limits.} We note that 
while the gamma-ray limits are severe for DM annihilation into two-body final states~\cite{DM1, bertone, bell, Cirelli:2009dv, Papucci:2009gd, Ackermann:2011wa, Hooper:2012sr, kopp, hooper}, the four-body final states (as in our case) are relatively less constrained~\cite{Papucci:2009gd}. In our model, photons can be emitted from external lepton legs, final-state radiation or in a subsequent decay of the pair of leptons directly produced from $\mathbf{\Delta}$ decay.  As can be seen from Fig.~\ref{fig3}, the NH case produces more photon flux compared to the IH case at higher energies. This is mainly due to the photons coming from tau decays in the NH case. The photon energy spectrum for the NH case agrees well  with the excess photon flux at high energies observed by Fermi-LAT, but eventually falls off to the background level as we approach the DM mass. More precise measurements of the photon flux at higher energies (beyond TeV) should be able to (in)validate this feature, and hence, provide a test of this model. 

Note that since the DM candidate in our model interacts with charged $SU(2)_L$-triplet scalars which directly couple to SM photons,  the DM can pair-annihilate to photons: $DD\to \gamma\gamma$ (and also $DD\to Z\gamma$) via one-loop diagrams mediated by charged scalars. This process can lead to a monochromatic gamma-ray line spectrum with energy  $E_\gamma=m_{\rm DM}$ for $DD\to \gamma\gamma$, and for $DD\to Z\gamma$, with photon-energy $E_\gamma=m_{\rm DM}(1-m_Z^2/4m_{\rm DM}^2)$. For a relatively light DM, this mechanism was invoked in~\cite{Wang:2012ts} to explain the 130 GeV $\gamma$-ray line spectrum feature recently reported~\cite{Bringmann:2012vr, Weniger:2012tx, Tempel:2012ey, Su:2012ft} in 
the publicly available Fermi-LAT data~\cite{Atwood:2009ez, Ackermann:2012kna}.  However, we find that our best-fit DM mass values explaining the AMS-02 positron excess are too high to account for the $E_\gamma=130$ GeV line signal, either directly from DM annihilation or from internal bremsstrahlung. Thus we conclude that it is very hard to simultaneously explain both the AMS-02 positron excess and the Fermi-LAT 130 GeV gamma-ray line signal in our model (as generically the case for any single-component DM model). 

\section{Effects of large $\lambda_\Phi$}\label{sec5}
\begin{figure}[t]
\centering
\includegraphics[width=6.5cm]{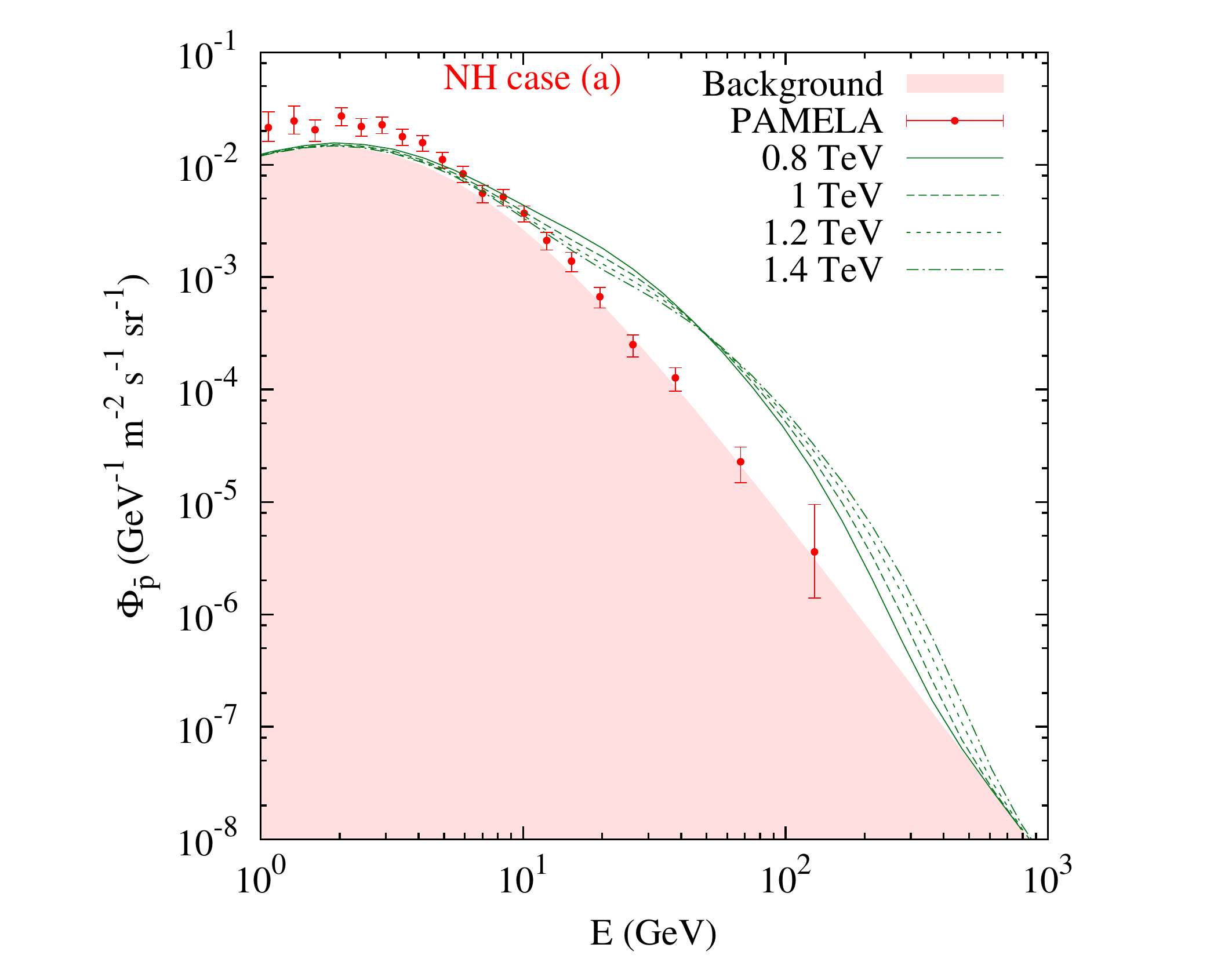}
\hspace{0.5cm}
\includegraphics[width=6.5cm]{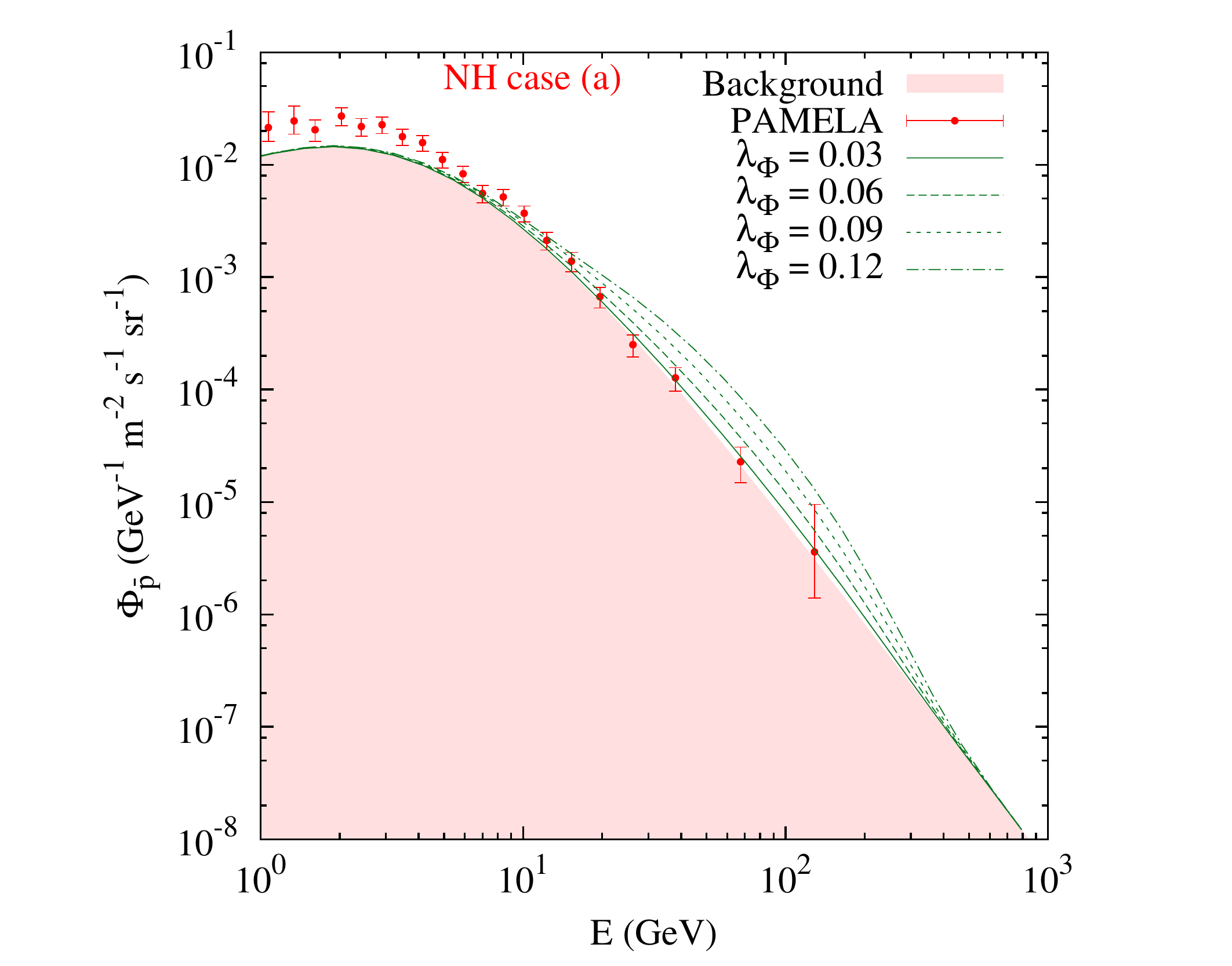}
\caption{The anti-proton flux in our model for a non-zero $\lambda_\Phi$ in the NH case (a). The left panel shows the variation with respect to different DM mass with a fixed $\lambda_\Phi=\lambda_\Delta=0.15$, whereas the right panel shows the variation with respect to different values to $\lambda_\Phi$ for a fixed DM mass $m_{\rm DM}=1$ TeV. The cosmic-ray anti-proton background (shaded region) as well as the PAMELA data (with error bars)~\cite{ap2} are also shown for comparison. The boost factors for different DM masses are of the same order as those for NH case (a) in Table~\ref{tab2}. }
\label{fig5}
\end{figure}

So far we have considered the case $\lambda^2_\Delta\gg \lambda^2_\Phi$ in Eq.~(\ref{an}). In this case, there is no significant excess in the anti-proton flux as the DM annihilation is dominantly to leptonic final states. In this section we justify this choice of parameters by noting that if we switch on a large $\lambda_\Phi$ coupling, the DM pair can annihilate to a SM Higgs pair which subsequently decays to mostly bottom quarks (58\% branching ratio for $m_h=125$ GeV) and $WW^*$ (22\% branching ratio). This has two adverse effects on the model: First, the hadronic final states will produce a sizable anti-proton flux. This is illustrated in Fig.~\ref{fig5} for $\lambda_\Phi=\lambda_\Delta$ and for NH. Also shown for comparison are the PAMELA data taken from~\cite{ap2}, and the expected background anti-proton flux from cosmic rays, obtained using {\tt GALPROP} with the same input parameters as given in Section~\ref{sec3} for the AMS-02 data fitting. It is clear that for any non-negligible value of $\lambda_\Phi$, the anti-proton excess will be inconsistent with the observed flux from PAMELA~\cite{ap2} which is in good agreement with the cosmic-ray anti-proton background. In fact, using this information, one can derive an upper limit on $\lambda_\Phi\lsim 0.06$, to be consistent with the PAMELA data, although , although this limit is subject to the propagation model uncertainties. A large $\lambda_\Phi$ will also lead to a large DM-nucleon spin-independent scattering cross section through a $t$-channel Higgs exchange. For instance,  for $m_h=125$ GeV, we have 
\begin{eqnarray}
\sigma_{\rm SI}\simeq (2\times 10^{-45}~{\rm cm}^2)\left(\frac{\lambda_\Phi^2}{0.1}\right)\left(\frac{1~{\rm TeV}}{m_{\rm DM}}\right)^2.
\end{eqnarray}
Thus an upper limit on $\lambda_\Phi\lsim 0.06$ from the anti-proton flux implies an upper limit on $\sigma_{\rm SI}\lsim 7\times 10^{-47}~{\rm cm}^2$ (for $m_{\rm DM}=1$ TeV) in our model. Note that this indirect limit derived from anti-proton flux constraints is about two orders of magnitude stronger than the current direct detection limit of $10^{-44}~{\rm cm}^2$ from LUX~\cite{Akerib:2013tjd} for a TeV-scale DM. 

\begin{figure}[t]
\centering
\includegraphics[width=6.5cm]{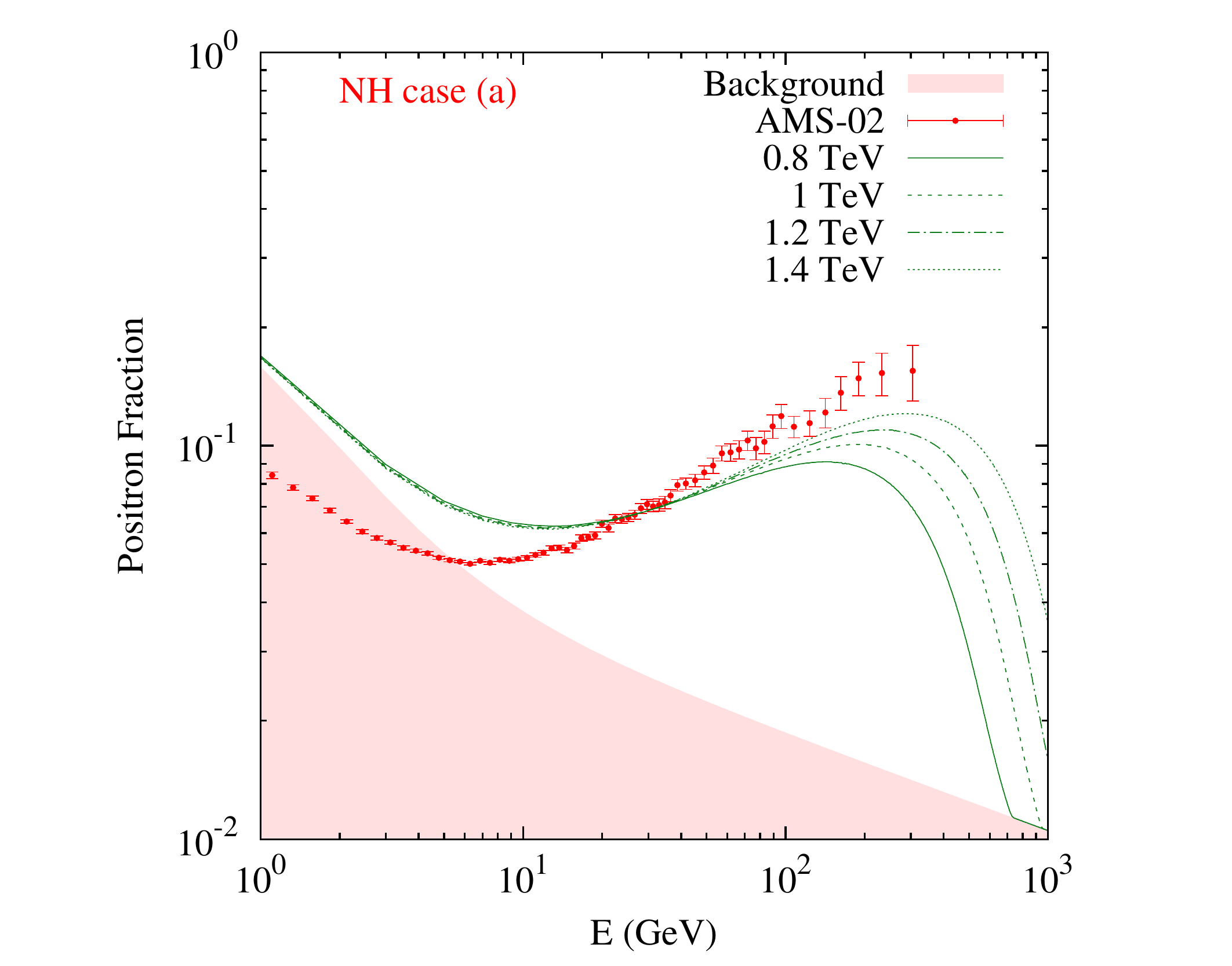}
\caption{The positron fraction as a function of the positron energy for different DM masses in our model for a large $\lambda_\Phi=\lambda_\Delta$. Also shown are the AMS-02 data~\cite{AMS} and the cosmic-ray background for comparison. The boost factors for different DM masses are of the same order as those for NH case (a) in Table~\ref{tab2}. This plot justifies our choice of $\lambda_\Delta\gg \lambda_\Phi$ in Sections~\ref{sec3} and \ref{sec4}.}
\label{fig6}
\end{figure}

Secondly, for a given DM mass the $b\bar{b}$ final states from the Higgs decay will lead to more positron flux in the lower energy regime due to the fact that the Higgs produced due to DM annihilation is highly boosted to start with, and moreover, the positrons coming from $b$-decay lead to a further widening of their energy spectrum. Thus, we would expect more positron fraction at lower energies than at higher energies in this case. This significantly worsens the fitting with AMS-02 data which has very precise fraction values in the low-energy bins. This effect is demonstrated in Fig.~\ref{fig6} for NH case (a) which had given us the best-fit scenario in Fig.~\ref{fig1}.  It is clear that this fit has a bad $\chi^2$/dof, e.g., 9.11 for $m_{\rm DM}=0.8$ TeV and 5.37 for $m_{\rm DM}=1.4$ TeV, because while trying to fit the low-energy AMS-02 data which now requires a smaller boost factor, we completely miss the data points in the high-energy bins. Due to these reasons, a large value of the Higgs quartic coupling $\lambda_\Phi$ is disfavored in this model.  
\section{Conclusion}\label{sec6}
We have discussed a simple extension of the SM by adding an electroweak scalar triplet field to take into account the non-zero neutrino masses by a low-scale Type-II seesaw mechanism, and in addition, a real scalar DM candidate satisfying the observed cold DM thermal relic density. This DM candidate has leptophilic annihilations for a wide range of the model parameter space, with the final-state lepton flavors intimately connected to the neutrino mass hierarchy. Using the recently released AMS-02 data on the positron fraction and flux, we find that a DM interpretation of the AMS-02 data in this model prefers a normal hierarchy of neutrino masses over an inverted hierarchy. The large boost factors required in the normal hierarchy case to explain the AMS-02 data are still consistent with the latest experimental limits derived from Fermi-LAT and IceCube data, except when the DM and triplet scalars are close to being mass-degenerate. The CMB constraints on the thermal DM annihilation rate can be avoided if the required boost factor has an astrophysical origin. We also find that the absence of an excess in the anti-proton flux as suggested by PAMELA requires the DM coupling to the SM Higgs boson in this model to be quite small. This in turn puts an indirect 
upper bound on the DM-nucleon scattering cross section which is about two orders of magnitude stronger than the current direct detection upper limit from LUX. With more precise data from the ongoing AMS-02, Fermi-LAT and IceCube experiments,  especially in the high-energy regime, in combination with the DM direct detection searches as well as future neutrino oscillation experiments capable of resolving the neutrino mass hierarchy, this simple model can be tested unambiguously. 
This information, in combination with the LHC searches for the doubly-charged 
scalars and more precise data on the Higgs-to-diphoton signal strength, 
might be able to completely probe the allowed parameter space of the minimal low-scale Type-II seesaw model as a single viable extension of the SM.         
\section*{Acknowledgments}
We thank M. Chakraborti, U. Chattopadhyay,  P. Majumdar,  A. Pukhov and S. Rao
for useful discussions. We also thank the anonymous referee for many constructive comments which helped 
improve the manuscript. DKG would like to acknowledge the hospitality provided 
by the University of Helsinki and the Helsinki Institute of Physics where part
of this work was done. NO would like to thank the Particle Theory Groups at the Maryland Center for Fundamental Physics, University of Maryland, and the Bartol Research Institute, University of Delaware, for hospitality during his visits.
The work of PSBD is supported by the 
Lancaster-Manchester-Sheffield Consortium for Fundamental Physics under STFC 
grant ST/J000418/1.  The work of NO is supported in part by the DOE Grant No. DE-FG02-10ER41714. 
 
\end{document}